\documentclass[oldversion]{aa}
\usepackage[pdftex]{graphicx}
\usepackage{epstopdf}
\usepackage{graphicx}
\usepackage{epsfig}
\usepackage{amsmath}
\usepackage{wasysym}

\usepackage{marvosym}
\usepackage{pxfonts}
\pdfoutput=1
\usepackage[pdftex]{color,xcolor}
\usepackage[pdftex]{graphicx}
\usepackage{epstopdf}
\usepackage{mathtools}
\usepackage{natbib}
\usepackage{breqn}
\usepackage{setspace}
\bibpunct{(}{)}{;}{a}{}{,}

\usepackage[]{hyperref}               % pour obtenir des liens hypertextes dans le pdf [draft]
\hypersetup{pdfauthor=Christoph Mordasini}
\hypersetup{pagebackref=true, breaklinks=true,colorlinks=true,urlcolor=blue, linkcolor=blue,  citecolor=blue,anchorcolor=blue,pagecolor=blue, bookmarks=true, bookmarksopen=true}

\def\mearth{M_\oplus}
\def\msun{M_\odot}

\def\mcore{M_{\rm core}}

\def\f1{f_{\rm I}}

\def\mstar{M_*}

\def\beq{\begin{equation}}
\def\eeq{\end{equation}}
\def\beqa{\begin{eqnarray}}
\def\eeqa{\end{eqnarray}}

\def\mearth{M_\oplus}
\def\rearth{R_\oplus}
\newcommand{\mz}{M_{\rm Z}}

\newcommand{\mdotxy}{\dot{M}_{\rm XY}}

\def\rj{R_{{\rm \jupiter}}}
\def\tint{T_{{\rm int}}}

\def\apjl{ApJL}                % Astrophysical Journal, Letters
\def\apjs{ApJS}               % Astrophysical Journal, Supplement
\def\ao{Appl.Optics}          % Applied Optics
\def\aap{A\&A}                % Astronomy and Astrophysics
\begin{document}

\title{Global Models of Planet Formation and Evolution}

\author{C. Mordasini\thanks{Reimar-L\"ust Fellow of the MPG}\inst{1}  \and P. Molli\`ere\inst{1} \and K.-M. Dittkrist\inst{1} \and S. Jin\inst{1,2}  \and Y. Alibert\inst{3,4}}

\institute{Max-Planck-Institut f\"ur Astronomie, K\"onigstuhl 17, D-69117 Heidelberg, Germany } 

\institute{Max-Planck-Institut f\"ur Astronomie, K\"onigstuhl 17, D-69117 Heidelberg, Germany \and
Purple Mountain Observatory, Chinese Academy of Sciences, Nanjing 210008, China \and
Center for space and habitability, Physikalisches Institut, University of Bern, Sidlerstrasse 5, CH-3012 Bern, Switzerland \and Institut UTINAM, CNRS-UMR 6213, Observatoire de Besan\c{c}on, BP 1615, 25010 Besan\c{c}on Cedex, France}

\offprints{Christoph Mordasini, \email{mordasini@mpia.de}}

\date{Received  ---  / Accepted ---}

\abstract
{Despite the strong increase in observational data on extrasolar planets, the processes that led to the formation of these planets are still not well understood. However, thanks to the high number of extrasolar planets that have been discovered, it is now possible to look at the planets as a population that puts statistical constraints on theoretical formation models. A method that uses these constraints is planetary population synthesis where synthetic planetary populations are generated and compared to the actual population.%}%context
%{
The key element of the population synthesis method is a global model of planet formation and evolution. These models directly predict observable planetary properties based on properties of the  {natal} protoplanetary disk, linking two important classes of astrophysical objects. To do so, global models build on the simplified results of many specialized  models that address one specific physical mechanism. %} %Goals
%{
We  {thoroughly} review the physics of the sub-models included in global formation models. The sub-models can be classified as models describing the protoplanetary disk (of gas and solids), those that describe one (proto)planet (its solid core, gaseous envelope, and atmosphere), and finally those that describe the interactions (orbital migration and N-body interaction).  {We compare the approaches taken in different global models,  discuss the links between specialized and global models, and identify physical  {processes} that  require improved  descriptions in future work.} % }
%{
We then shortly address important results of planetary population synthesis like the planetary mass function or the mass-radius relationship. With these statistical results, the global effects of  physical mechanisms occurring during planet formation and evolution become apparent, and specialized models describing them can be put to the observational test.%}% Results
%{
 Due to their nature as meta models,  global models depend on the results of  specialized models, and therefore on the development of the  field of planet formation theory as a whole. Because there are important uncertainties in this theory, it is likely that the global models will in future undergo significant modifications. Despite these limitations, global models can already now yield many testable predictions. With future global models addressing the geophysical characteristics of the synthetic planets, it should eventually become possible to make predictions about the habitability of planets based on their formation and evolution.} %Conclusions. 

\titlerunning{Global Models of Planet Formation and Evolution}
\authorrunning{C. Mordasini et al.}

\maketitle
\section{Introduction}\label{sect:intro}
Thanks to the progress of observational techniques in the last decades, we are the first generation of human beings that has had the technological capabilities to answer the question about the existence of planets around other stars \citep{mayorqueloz1995}. Since then, there was an enormous increase in observational data on extrasolar planets. The latest observational results from different detection techniques \citep{mayormarmier2011,boruckikoch2011b,cassankubas2012} even indicate that the presence of planets  is the rule rather than the exception, at least around solar-like stars. However, this  increase in observational data on extrasolar planets does  not mean that we can now fully explain  how these planets came into existence from a theoretical point of view. On the contrary, many observational findings on extrasolar planets were not predicted from planet formation and evolution theory, or were even in contrast to it, showing that this field is still in its infancy.  

A young method to improve the theoretical understanding of planet formation is planetary population synthesis. It is a statistical method that makes it possible to improve the theoretical understanding of the  physics governing  planet formation and evolution by using statistical comparisons to  observational constraints provided by the population of extrasolar planets.  With this approach the global effects of many key physical  {processes} occurring during planet formation and evolution can be put to the observational test, something which is notoriously difficult in astronomy, since the objects that are studied are far away and only accessible via the radiation they (or their host star) emit.

In this paper we review global models of planet formation and evolution that are used in such planetary population synthesis calculations \citep{idalin2004a,idalin2004,idalin2005a,idalin2008,idalin2008c,thommesmatsumura2008a,mordasinialibert2009a,mordasinialibert2009,miguelbrunini2009,idalin2010,alibertmordasini2011a,mordasinialibert2012d,mordasinialibert2012a,mordasinialibert2012,hellarynelson2012,alibertcarron2013,forganrice2013,hasegawapudritz2013,idalin2013,galvagnimayer2013,benzida2013}. In this context, ``global'' means that these models can directly predict planetary properties based on properties of the protoplanetary disk in which the planets form. To do so, they unite in one model the results of many specialized  models that address  a specific important mechanism occurring during planet formation like accretion or migration.  Here we  concentrate on the physical description of these  {processes} as included in the global models because this is the key ingredient of the entire population synthesis approach. 

\subsection{Observational motivation}
Currently, there are  approximately  {1500} confirmed exoplanets known\footnote{Regularly updated  databases  can be found at  www.exoplanet.eu \citep{schneiderdedieu2011} and www.exoplanets.org \citep{wrightfakhouri2011}.}  that were mostly  {detected with} the spectroscopic radial velocity technique  {or photometric transit observations}. Additionally, there are about  {four} thousand candidates from the \textsc{Kepler} satellite \citep{boruckikoch2011b}  that were detected with extremely precise  transit photometry. These detections have revealed an exiting diversity in the properties of planetary companions that was not expected from the structure of our own planetary system, the Solar System. The detections have, however, not only revealed a surprising diversity, but also a number of interesting correlations and structures in the properties of the planets.  

These insights were in particular possible thanks to the large number of  planets now known. For the first time, this allows  to look at the extrasolar planets no more  solely as  single objects.  Instead, it  is possible to look at them as a population that is characterized by a number of statistical properties. Important examples are the distributions of masses, semimajor axes, radii, eccentricities and the relations between theses quantities. Understanding these statistical properties from the point of view of planet formation and evolution theory is one of the fundamental goals of the models presented here. The understanding that can be gained in this way  also feeds back into the way we understand our own Solar System. Also the Solar System itself provides a large body of precise observational constraints against which planet formation models must be compared. But developing a theory that is focused on one planetary system can be misleading; the discovery of exoplanets that are very different from any Solar System planet has shown that.  

The special interest in a statistical population-wide approach also comes from the fact that the knowledge about a single extrasolar planet is often limited. For the large majority of the extrasolar planets, still only a few orbital elements (semimajor axis, eccentricity,...) and a minimum mass are known (or a radius, but no mass in the case of most \textsc{Kepler} candidates). In order to  benefit also from the large number, but individually limited data sets, statistical methods are necessary. Having this ability is important since several future surveys like the \textsc{Gaia} space mission or the \textsc{Sphere} and \textsc{Gpi} direct imaging surveys \citep{beuzitfeldt2008,mcbridegraham2011} will yield additional statistical data sets. 

For a handful planets this has partially changed  in the last few years, and a first  rough geophysical characterization of some extrasolar planets has become possible by multi-band photometry or spectroscopy \citep[e.g.,][]{richardsondeming2007,konopackybarman2013}. These are typically planets around bright stars for which both the mass and the radius are known (or alternatively the intrinsic luminosity in the case of directly imaged planets). In this case,  more observational constraints can be derived like the mean density or the atmospheric structure and composition. These planets are investigated in  detail, and will likely have a special role in the next decade of extrasolar planet study. This observational progress was the motivation for some recent work on the theoretical side discussed in this paper, which is the extension of an existing planet formation model \citep{alibertmordasini2005} into a self-consistently coupled planet formation and evolution model \citep{mordasinialibert2012d}. With such a combined model one can predict all major observable characteristics of a planet. We will discuss this topic and how it can be incorporated into statistical studies in Sect. \ref{sect:fromformationtoevolution}. 

The fundamental assumption behind the  planetary population synthesis method is that the observed statistical properties of exoplanets (like the aforementioned distributions) can be explained by the action of the always identical governing physical processes during the formation phase of the planets, but under different  initial conditions. The initial conditions for the planet formation process are the properties of the protoplanetary disk which are found to surround most newly born stars \citep[e.g.,][]{haischlada2001,fedeleancker2010}. Observations of such circumstellar disks show that they also have, as planets, a wide variety of properties in terms of their most important characteristics like their mass, lifetime, or radius \citep[e.g.,][]{andrewswilner2010}. For an individual planetary system, the properties of the protoplanetary disk from which it formed are mostly unknown, except maybe for the dust-to-gas ratio in the disk that is likely correlated with the stellar metallicity that can be measured spectroscopically today. This means that the initial conditions are only known in a statistical sense, which again makes a statistical approach appropriate. It is clear that this fundamental assumptions neglects that stars, and therefore planets, usually do not form in isolation, but in stellar clusters. This environment can influence the formation process. Two examples are the impact of close-by massive stars on the lifetime of a protoplanetary disk \citep{adamshollenbach2004}, or the gravitational perturbation by a passing star or temporary binary companion \citep{malmbergdavies2010}.

\section{Observational constraints}
In comparison to  other studies in the domain of planet formation theory, population synthesis is directly at the interface between theory and observation, since it is  a goal to connect these two domains. It is therefore interesting to  briefly discuss two central observational results that are important for the theoretical studies, namely, the semimajor axis - mass and the mass - radius diagram. 

It is clear that besides these two results, there is a very large number of additional statistical constraints that can be deduced from the extrasolar planets like the distributions of radii, eccentricities, and luminosities,  {as well as} the relations between all these quantities. Further important observational constraints are the frequency and properties of planets as a function of  host star properties like mass and metallicity, the mean spacing between planets in multiple systems, the alignment of the planets among themselves and with the stellar equator, the frequency of planets in mean motion resonances and so on.  {A recent review on the observed properties of extrasolar planets can be found in \citet{fischerhoward2014}.}

\subsection{Semimajor axis - mass diagram}
\begin{figure}
\begin{center}
       \includegraphics[width=\columnwidth]{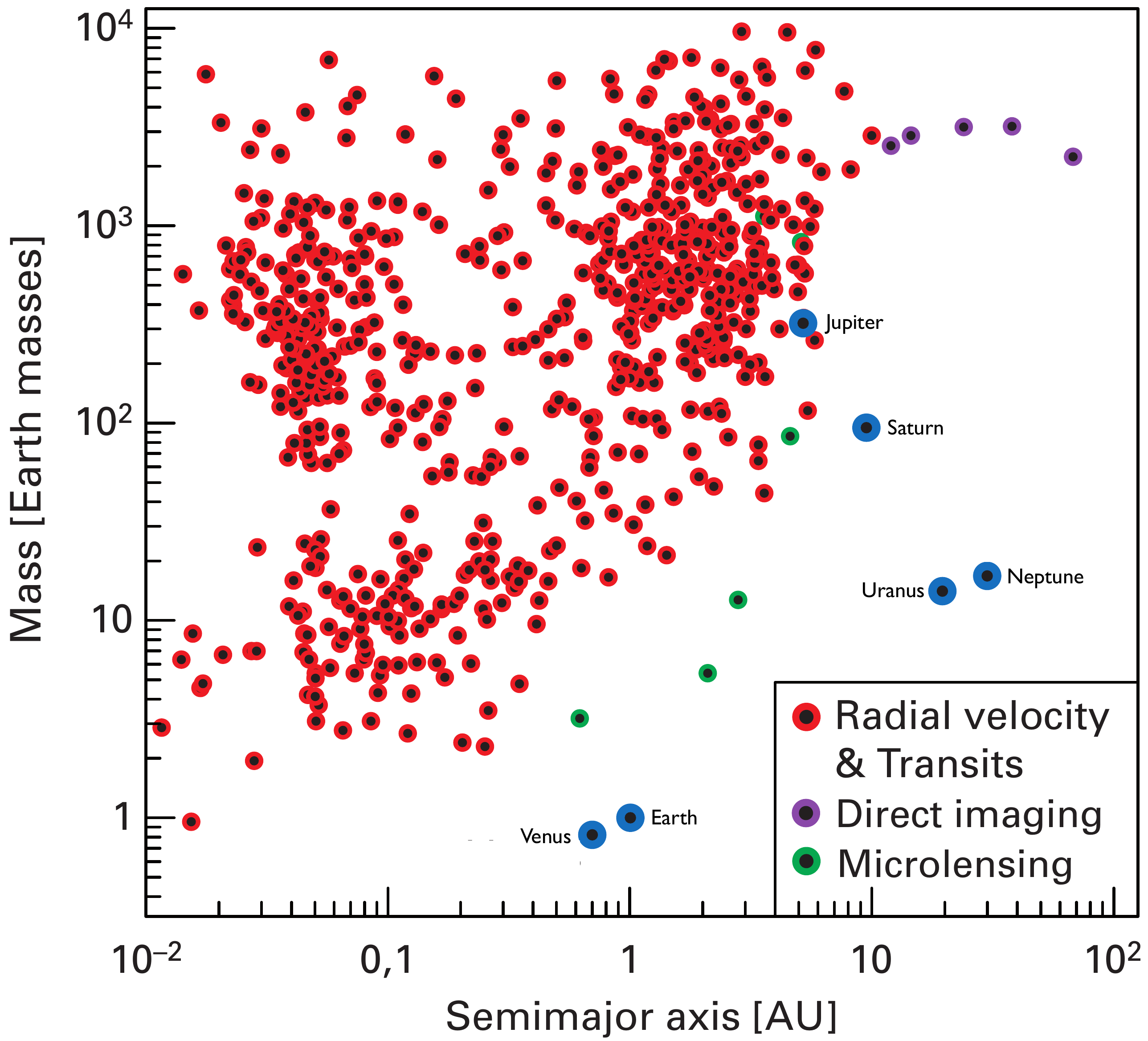}
\caption{Semimajor axis - mass diagram of  extrasolar planets. The  colors show the observational  technique that was used for the detection. The figure is not corrected for the various observational biases that favor for the radial velocity and the transit technique the detection of close-in, giant planets. The planets of the Solar System are also shown for comparison.  {Data from www.exoplanet.eu \citep{schneiderdedieu2011}.}}\label{fig:INTROam}
\end{center}
\end{figure} 

Figure \ref{fig:INTROam} shows the planetary semimajor axis-mass ($a$-$M$) diagram. It is a classical observational constraint for population synthesis and is still one of the most important observational results. Explaining the structures seen in this plots is one of the goals of planetary population synthesis. The extreme diversity, but also the existence of certain structures in the $a$-$M$ diagram is  {visible}. For extrasolar planets, the mass-distance diagram has become a representation of similar importance as the Hertzsprung-Russell diagram for stellar astrophysics \citep{idalin2004a}. 

In the plot, one can distinguish several groups of planets. There are, for example, massive close-in planets without an equivalent in the Solar System. Such hot Jupiters  are found around approximately 1\% of solar-like stars \citep{marcybutler2005,howardmarcy2010,mayormarmier2011}. A class of extrasolar planets that has only been detected in the last few years thanks to the progress in the observational precision are low-mass planets with masses between 1 to 30 $\mearth$ (Earth masses). These super-Earths and mini-Neptunes seem to be very abundant, since every second FGK star is found to have such a companion with a period of up to 100 days \citep{mayormarmier2011}.  This result is at least in qualitative agreement with the analysis derived from the \textsc{Kepler} mission which also detects an extremely numerous population of planets with small radii $\lesssim 4 \rearth$ \citep[e.g.,][]{howardmarcy2012,fressintorres2013a}. Since hot Jupiters are much more easily detected by both the radial velocity and the transit method compared to low-mass (respectively small) planets, the number of such low-mass planet is underestimated in Fig. \ref{fig:INTROam} which is not corrected for  observational biases. 

If we further inspect Figure \ref{fig:INTROam}, we may ask whether it points to a statistically significant deficit of planets with a mass of approximately 40 Earth masses \citep[so-called ``planetary desert'',][]{idalin2004}. This is, among many others, a very interesting question from a theoretical point of view that will be discussed later on \citep[Sect. \ref{sect:INTROrvcomp}, see also][]{mordasinimayor2011}. 

The semimajor axis distribution and the planetary mass function are two fundamental 1D statistical distributions that are encoded into the $a$-$M$ diagram.  These distributions are (besides of the radius distribution) of prime interest for statistical studies \citep[e.g.,][]{idalin2004,thommesmatsumura2008a,mordasinialibert2009a} and can be compared to theoretical results  with Kolmogorov-Smirnov tests  \citep{mordasinialibert2009}. The planetary initial mass distribution is further addressed  in Sect. \ref{sect:INTROrvcomp}. 

In the figure there are also planets that were discovered by direct imaging. In this technique, one measures of course not the mass, but the luminosity of a planet. The conversion of luminosity into mass is model dependent and uncertain as shown by \citet{marleyfortney2007} or \citet{spiegelburrows2012}. In \citet{mordasini2013} a new aspect was  pointed out  which is important for the conversion of luminosity into mass. It is found, perhaps surprisingly at first sight, that the post-formation luminosity of giant planets formed by core accretion depends significantly on the mass of the solid core. We  address this finding in Sect. \ref{sect:INTROdirectimaging}.

\subsection{Mass-radius diagram}\label{sect:obsmassradiusdia}
\begin{figure}
\begin{center}
\includegraphics[width=1\columnwidth]{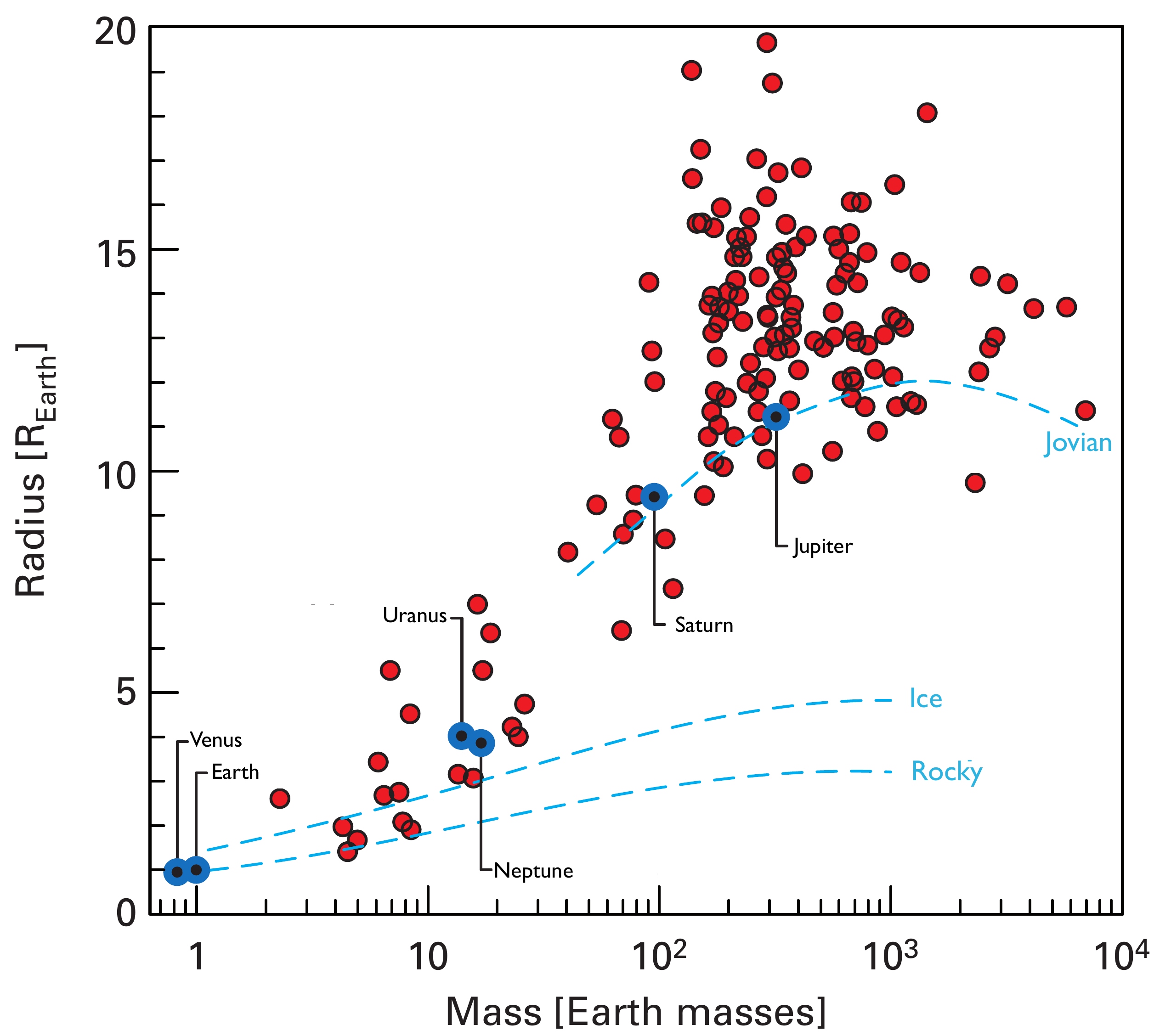}
\caption{The observed mass-radius relationship of the extrasolar planets (red points) together with theoretical mass-radius relationships for planets with an Earth-like composition, with an interior consisting purely of water ice, and for planets with a bulk composition roughly like Jupiter. The planets of the Solar System are also shown.  {Data from www.exoplanets.org \citep{wrightfakhouri2011}.}}\label{fig:INTROmrobs}
\end{center}
\end{figure}

Figure \ref{fig:INTROmrobs} shows the observed mass-radius diagram of the extrasolar planets  and compares it with three theoretical mass-radius relationships for planets with different bulk compositions \citep[from][]{mordasinialibert2012}.  The combination of measurements of the radius of a transiting planet (first by \citealt{henrymarcy2000} and \citealt{charbonneaubrown2000}) and its mass (via radial velocity) makes it possible to derive the mean density of the planet. In the past years, such combined measurements were made for many exoplanets, so that the planetary mass-radius diagram became known\footnote{Note that the mass-radius relation is a function of time at least for planets with a significant gaseous envelope because they contract on Gyr timescales. The $a$-$M$ is usually more static at late times (several Gyrs after formation). But at early times (typically a few 10-100 Myrs after the dissipation of the protoplanetary disk) it also evolves due to giant impacts, gravitational interactions, and atmospheric mass loss. }.  It is an observational result of similar importance as the semimajor axis-mass diagram. One notes that as in the $a$-$M$ diagram there is a large diversity, but that there are also clear trends leading, e.g.,  to regions in the plot that are not populated. 

A surprising observational result was the discovery of numerous ``inflated'' planets with radii much larger than Jupiter  {which} was not predicted from standard planet evolution theory. It is now clear \citep{millerfortney2011,demoryseager2011a} that these bloated radii are related to the proximity of the planets to the host star (most currently known transiting planets have small orbital distances of a few 0.1 AU or less due to the decreasing geometrical transit probability with distance). The exact mechanism that leads to the large radii is still not completely understood. Several possible explanations have been put forward in the past, including tidal heating \citep{bodenheimerlaughlin2003}, dissipation of stellar irradiation deep in the atmosphere \citep{guillotshowman2002}, double diffusive convection \citep{chabrierbaraffe2007}, enhanced atmospheric opacities \citep{burrowshubeny2007},  and ohmic dissipation \citep{batyginstevenson2011}. 

The importance of the $M$-$R$ diagram stems from its information content about the inner bulk composition of planets which is the first very basic geophysical characterization of a planet. This first  characterization is found by the comparison of the observed mass and radius with theoretical models of the internal structure \citep[e.g.,][]{fortneymarley2007a,seagerkuchner2007}. In the Solar System, there are three fundamental types of planets, namely, terrestrial, ice giant and gas giant planets. The imprint of the bulk composition on the radius is indicated by three theoretical lines in Figure \ref{fig:INTROmrobs}. Two lines show the theoretical mass-radius relationship for solid planets made of silicates and iron in a 2:1 ratio as for the Earth and for pure water planets, while the third line shows the $M$-$R$ relationship for giant planets consisting  of H/He and a solid core of about 10\% in mass. Being able to understand and to reproduce this second fundamental figure besides the $a$-$M$ diagram is another goal of statistical studies of planet formation and evolution. The reason for the importance  of the $M$-$R$ diagram for formation theory are  the additional constraints on the formation process that cannot  be derived from  the mass-distance diagram alone. 

An example are the observational constraints coming from the $M$-$R$ diagram on the radial extent of orbital migration. Efficient inward migration brings ice-dominated, low-density planets from the outer parts of the disk close to the star. These planets can in principle be distinguished from denser planets consisting only of silicates and iron that have presumably formed more or less  in situ in the inner  hotter parts of the disk. Complications arise from the fact that the mass-radius relationship is degenerate in some parameter space (different bulk compositions can lead to an identical mass-radius relationship, see \citealt{rogersseager2010}). Therefore, spectroscopic measurements might be necessary to actually distinguish the two types of planets. Other key questions are: 1. What are the heavy element masses contained in  giant planets? This is related to the question about the fundamental formation mechanism of giant planets, core accretion or gravitational instability. 2. Which planets can accrete and also keep primordial H/He envelopes (envelope evaporation)? Figure \ref{fig:INTROmrobs} shows that there are low-mass planets which likely contain important amounts of H/He. They therefore form a class of low-mass, low-density planets without counterpart in the Solar System. 3.  Are there correlations between the planetary bulk composition and stellar properties (metallicity) as has been found by \citet{guillotsantos2006} or  \citet{burrowshubeny2007}?

\section{From detection to characterization}\label{sect:fromformationtoevolution} 
The above three questions are an observational motivation to extend a global planet formation model into a coupled formation and evolution model as in \citet{mordasinialibert2012a}. It then becomes possible to calculate radii based on the bulk composition obtained during formation, which makes new observational constraints usable for population synthesis. Such a model can then be used to study the population-wide mass-radius relationship and to compare with the radius distribution found by the \textsc{Kepler} satellite \citep[][]{mordasinialibert2012}. Compared to other, well-established planet evolution models \citep[e.g.,][]{burrowsmarley1997,baraffechabrier2003}, the evolutionary model of \citet{mordasinialibert2012a} is, however, still  significantly simplified in several aspects as discussed in Sect. \ref{sect:INTROatmopla}.  In view of future observations yielding very precise radii (e.g., by the photometric CHaracterizing ExOPlanet Satellite \textsc{Cheops}, \citealt{broegfortier2013}) it will probably be necessary to find more accurate physical descriptions also in global models.  

The mass-radius diagram represents, in a prototypical way, the transition of the focus from pure exoplanet detection to beginning exoplanet characterization in the past few years. Besides the $M$-$R$ relationship, there was recent observational progress towards characterization in two other domains:

\subsection{Direct imaging}
The first technique besides transits that has recently yielded important new  results for planet characterization is the direct imaging technique.  The method is technically challenging due to the small angular separation of a very faint source (the planet) from a much brighter one (the host star). The number of planets detected by direct imaging is currently still low. But already these discoveries, like the planets around HR 8799 \citep{maroismacintosh2008} or $\beta$ Pictoris \citep{lagrangebonnefoy2010} have triggered numerous theoretical studies regarding their formation \citep[e.g.,][]{dodson-robinsonveras2009,krattermurray-clay2010}. Two points about these planets are interesting: their large semi-major axis and the fact that we directly measure the intrinsic luminosity at young ages in several  IR bands. Both quantities are important to understand the formation mechanism (core accretion or gravitational instability) and in particular the physics of the accretion shock occurring when the accreting gas hits the planet's surface during formation \citep[e.g.,][]{commerconaudit2011b}. If the gravitational potential energy of the accreting gas is radiated away, low entropy gas is incorporated into the planet, leading to  a faint luminosity and small radius (so-called ``cold start'', \citealt{marleyfortney2007}) while the accretion of high entropy material leads to a ``hot start'' with a high luminosity and large radius \citep[e.g.,][]{burrowsmarley1997,baraffechabrier2003}. Recently, \citet{spiegelburrows2012} have shown that the different scenarios result in observable difference in the magnitudes of the young planets.  The global model mainly discussed in this work (see Fig. \ref{fig:INTROschema}) calculates the luminosity during both the formation and evolution phase with a self-consistent coupling. For young giant planets, this is a significant  {difference} compared to purely evolutionary models, since this coupling is necessary to know the entropy in the envelope directly after formation and to correctly predict the luminosity at young ages.  Since multi-band photometry can be used to estimate the metal enrichment of a planet and because new direct imaging instruments are  {currently} becoming operational  (\textsc{Sphere}  {and} \textsc{Gpi}), it is important that future global models will   {include better descriptions} of the gas accretion shock and better atmospheric models (cf. Section \ref{sect:INTROatmopla}).

\subsection{Spectroscopy}
Second, one of the most important aspects of the recent observational progress towards characterization are the spectra of a number of exoplanets transiting bright stars \citep[e.g.,][]{richardsondeming2007}. The atmosphere represents a window into the composition of a planet and contains a multitude of clues to its formation history. The atmospheric composition depends on the composition of the host star, the nebula properties like the temperature where the planet formed, the composition of the accreted gas and planetesimals, the size of the planetesimals and their (material) strength, the evolution of the distribution of the chemical compounds inside the planet, and so on. Each migration and accretion history will result in a different atmospheric composition, as well as core and total heavy element mass. Jupiter, for example, is enriched in carbon by about a factor three relative to the sun, while Uranus and Neptune are enriched by a factor $\sim$30 \citep[e.g.,][]{guillot1999}. An enrichment relative to the sun is a natural prediction of the core accretion formation model  but not of the competing gravitational instability model.  {For example,  planet formation simulations based on the core accretion paradigm that reproduce (some of) the observed chemical composition like the enrichment in carbon were presented in \citet{gautierhersant2001} and \citet{alibertmousis2005}. One should however note that other aspects of the composition of Jupiter are not straightforward to understand neither in the context of the core accretion nor gravitational instability model. In particular, the measurement that the enrichment of Jupiter both in  highly volatile argon on the one hand and the more refractory sulphur on the other is similar (again by a factor of 2-4 relative to solar, \citealt{owenmahaffy1999}) is difficult to explain in the context of conventional models of trapping of highly volatile gases in amorphous ice and a growth of Jupiter in a region similar to its current position. Several explanations  to this conundrum have been put forward like a formation of  the planetesimals that enriched Jupiter's envelope at lower temperatures and thus probably at significantly larger orbital distances \citep{owenmahaffy1999} or the incorporation of noble gases in the form of clathrate hydrates in  crystalline ice \citep{gautierhersant2001}.  } 

An exoplanet that lately got a lot of attention as an example how spectroscopy constrains planet formation and evolution theory is GJ1214b \citep{charbonneauberta2009}. It has a radius of about 2.7 Earth radii and a mass of 6.5 Earth masses, so that it is a planet without counterpart in the Solar System between terrestrial planets and ice giants. The measured mass and radius  {is compatible with} different internal compositions, like a rocky core with a hydrogen/helium atmosphere or a planet dominated by water with a water vapor atmosphere \citep{rogersseager2010}. Such different compositions, and the associated different mean molecular weight lead to different transmission spectra \citep[e.g.,][]{beandesert2011}.  {Initially} observed spectra showed that GJ1214b must either have an atmosphere with high-altitude clouds or hazes, or contain at least 70\% water by mass \citep{beandesert2011,bertacharbonneau2012}.   {Recent precise} observations could rule out cloud-free atmospheres even of high mean molecular mass \citep{kreidbergbean2014}. A clear solar-composition atmosphere is excluded with very high confidence. Combined formation and evolution simulations that keep track of where planetesimals deposit their mass during impacts in the gaseous envelope \citep{mordasinialibert2006} indicate that  highly enriched atmospheres are a typical outcome for a planet like GJ1214b with clear consequences for spectroscopy \citep{fortneymordasini2013}. But in general, despite the fact that some processes have been identified \citep{obergmurray2011,madhumousis2011,thiabaudmarbeuf2013}, no self-consistently linked calculations have been made to date that keep track of the chemical composition of both the accreted gas and planetesimals during formation and directly predict the atmospheric composition and spectrum. 

\section{Elements of the population synthesis method}\label{sect:INTROstatmeth}

\begin{figure*}
\begin{center}  
{\includegraphics[width=0.8\textwidth]{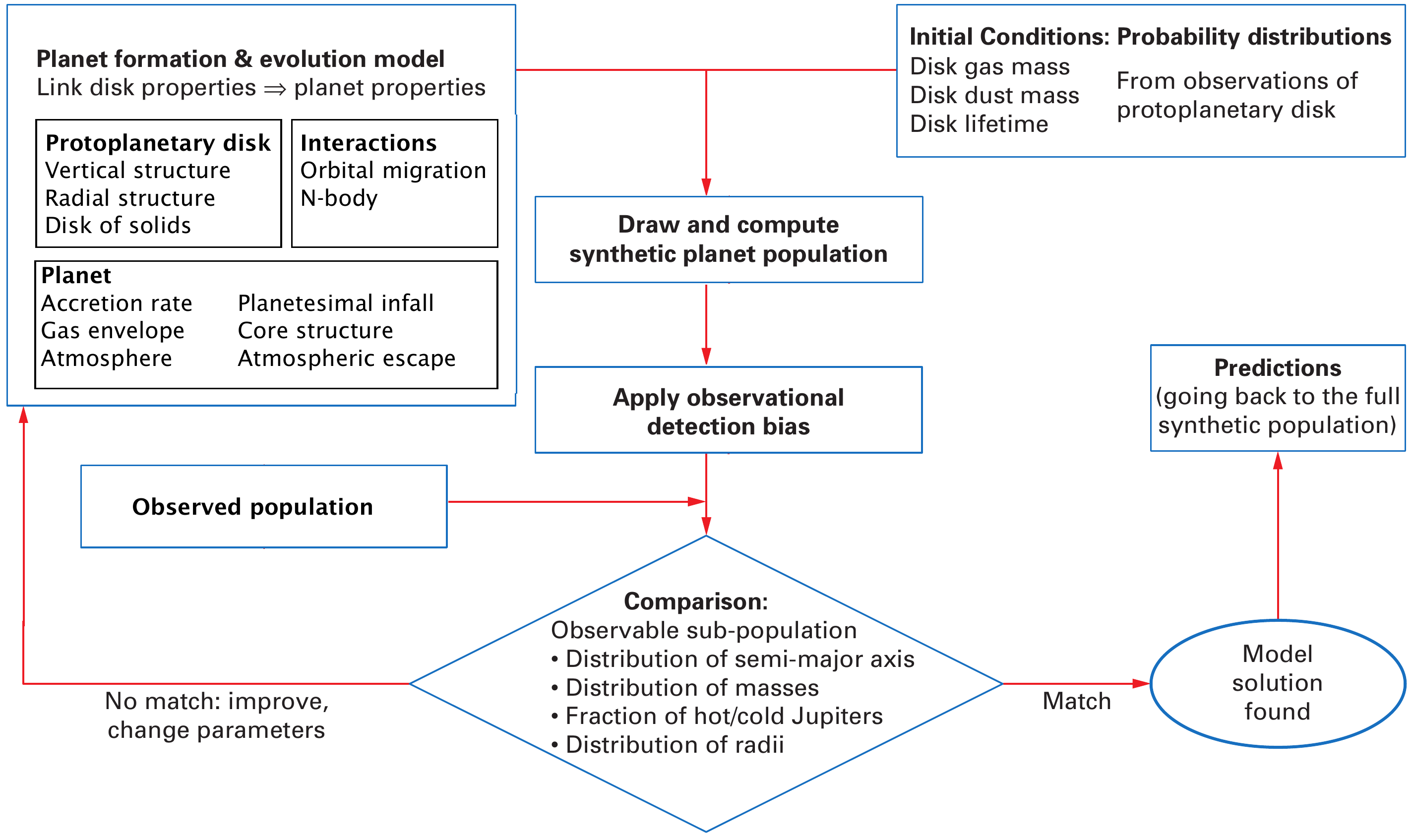}}
\caption{Schematic representation of  {workflow in the population synthesis method \citep{idalin2004a,mordasinialibert2009a}. The} eleven computational sub-models of  {the} combined global planet formation and evolution model  {are} based on the core accretion paradigm  {(see \citealt{alibertmordasini2005}, \citealt[][]{mordasinialibert2012a, mordasinialibert2012}, \citealt{alibertcarron2013})}.} \label{fig:INTROschema} 
\end{center}
\end{figure*}

Planetary population synthesis as a suitable method for statistical studies of planet formation and evolution was introduced in the pioneering work of \citet{idalin2004a}. A similar framework, but intended for more quantitative comparisons was established in \citet{mordasinialibert2009a}. The basic idea is to run a global planet formation model for varying initial conditions. With this framework the population-wide, statistical consequences of a theoretical description of a specific physical mechanism can be studied and compared with the population of actual extrasolar planets. Examples are specialized models of type I migration or of grain growth in protoplanetary atmospheres, see Section \ref{sect:transits}. The possibility to test specialized models is an important aspect of population synthesis. These specialized models are typically more complex and contain more subtleties than their simplified counterpart embedded in a global model. But if the simplified counterpart is still able to capture the essence of the original specialized model, then population synthesis is often the only possibility to test them observationally. A framework for population synthesis typically consists of the following elements  {that are shown in the flowchart of the method in Figure \ref{fig:INTROschema}. }

\subsection{Global planet formation and evolution model}
The most important element of population synthesis is a global planet formation and evolution model that establishes the link between disk and planetary properties. The sub-models of the global models typically used in population synthesis are described below in Section \ref{sect:INTROglobalmodel} and are the main subject of this paper.  For the different sub-models, already relatively well established standard physical descriptions are employed if possible, which are the result of specialized models.  Important simplifications are, however, often necessary  for computational time restrictions. It is clear that the current global models (and often also the specialized ones) only provide a first, very rough approximation of the complex  {processes} that  actually govern planet formation. In this sense it is likely that the global models will in future undergo important modifications, tracing in this way the developments in the field of planet formation theory.  In order to still test the global models as far as possible, dedicated simulations are made for relatively well known individual planetary systems, in particular the Solar System. But also some extrasolar systems can be studied individually, like, for example, the planetary system around HD 69830  with three Neptunian planets \citep{lovismayor2006,alibertbaraffe2006}.

A global formation models should  output as many observable quantities as possible, since in this case, one can use combined constraints from many techniques. The typical outputs are: the planetary mass, orbital distance and eccentricity (for comparison with radial velocity and microlensing), the radii for comparisons with transit observations, the intrinsic luminosity for comparison with discoveries made with direct imaging, and the atmospheric structure and composition for comparison with spectroscopy. The ability to compare concurrently and self-consistently with many different observational techniques (and therefore different sub-populations of planets) is crucial because global models suffer from the fact that they necessarily rely on a relatively high number of ill-constrained parameters (e.g., the viscosity parameter $\alpha$, the sizes of the planetesimals, the initial radial slope of the solids etc.). The more comparisons are possible, the better these quantities can be determined, and the less likely it becomes that agreement of the model with observations is only obtained because of a sufficiently high number of  unconstrained quantities.

\subsection{Probability distributions for the initial conditions}
The second central ingredient for population synthesis are sets of initial conditions. These sets of initial conditions are drawn in a Monte Carlo way from probability distributions. These probability distributions represent the different properties of protoplanetary disks and are derived as closely as possible from results of disk observations. At least three different fundamental disk properties have been considered in past population synthesis studies: The total disk (gas) mass \citep{beckwith1996,andrewswilner2010}, the dust-to-gas ratio  (assumed to be correlated with the stellar [Fe/H], \citealt{santosisraelian2004,fischervalenti2005}), and the lifetime of the disk \citep{haischlada2001,mamajek2009,fedeleancker2010}. Additionally properties can be the outer radius of the disk (controlled by the angular momentum of the collapsing cloud) or the initial radial slope of the solid surface density \citep{kornetstepinski2001,miguelguilera2011a}.  {It is important to note that the derivation of such distributions for the initial conditions is not straightforward, introducing uncertainties in the final populations. For example, the total disk mass is typically found from sub-mm observations of cold dust at large semimajor axes \citep[e.g.,][]{andrewswilliams2007,andrewswilner2010} to which gas is added at a ratio that is typical for the interstellar medium (usually a factor 100).  The disk masses that are derived with this method can be up to one order of magnitude smaller  compared to disk mass estimates based on the stellar accretion rate \citep{hartmanncalvet1998}. This discrepancy can result from a substantial growth of the dust to sizes to which  submm observations are no more sensitive \citep[e.g.,][]{andrewswilliams2007}. Furthermore, the concept of an ``initial'' protoplanetary disk  is in any case questionable since disks form dynamically during the infall of the protostellar core. A more realistic model for the initial conditions would therefore include the early formation phase of the disks based on a simple infall model \citep[e.g.,][]{huesoguillot2005}.   }

\subsection{Synthetic detection bias}
For a given set of initial condition, the formation model is used to calculate the final outcome, i.e., the planetary system. This step is repeated many times  leading to a population of synthetic planets (typically  $\sim$$10\,000$ planets). Many of these synthetic planets could not be detected by current observational techniques because their mass (or radius) is too small. In order to make quantitative comparisons with the observations, one must therefore apply a synthetic  observational detection bias. This leads to the sub-population of detectable synthetic planets. This group is then compared with  a comparison sample of actual exoplanets. Depending on the observational technique, different detection biases are used, meaning that typically, different sub-populations are probed.  It is clear that for quantitative comparisons, the selection bias of a given observational survey should be known as accurately as possible for this step. This makes that large, well characterized surveys like, e.g., the \textsc{Harps} high-precision survey \citep[][]{mayormarmier2011}, the \textsc{Kepler} satellite \citep{boruckikoch2011b}, and also microlensing surveys \citep[e.g.,][]{goulddong2010,cassankubas2012} are of particular interest.  

\subsection{Comparisons/Statistical tests}
For the comparison of the detectable synthetic planets and the actual planets various statistical methods can be used, like, for example, 2D Kolmogorov-Smirnov tests in the $a$-$M$ or $M$-$R$ plane. Other quantities that are tested are the detection frequency, and all different 1D distributions. It can further be studied if correlations exist between the initial conditions and the planet properties \citep{idalin2004,mordasinialibert2012d}, and if similar correlation exist in reality. The most important observed correlation is the one between the stellar metallicity and the frequency of giant planets. Giant planets are much more frequent around high metallicity stars \citep{gonzalez1997,santosisraelian2004,fischervalenti2005}, a correlation that can be reproduced with formation models based on the core-accretion theory \citep{idalin2004,mordasinialibert2009}. 

Depending on the results of this procedure, one can judge if the theoretical model is able to reproduce certain observed properties. In the ideal case, one single population should reproduce all observational constraints coming from many different techniques in a statistically significant fashion. In reality, there will always be differences between the model and the observations. This is, however, not an issue but instead the modus operandi of population synthesis, because the reasons for these differences are then analyzed, so that various physical descriptions of the  {processes} occurring during  formation and evolution can be tested. This can have the consequence that  an existing sub-model must  be modified or even abandoned for being inconsistent with observations,  or that new physical mechanisms must be added to the theoretical model. This is the fundamental process by which statistical studies improve the understanding of planet formation and evolution. Clearly, we currently still stand  at the beginning of this process, even if the global formation models discussed below already concentrate a non-negligible amount of physics in one  {framework}.  

\subsection{Predictions}
In case of a satisfactory agreement between theory and observation (at least for a given aspect), one can return to the full underlying synthetic population and make predictions about planets or planetary properties that currently cannot be observed yet, like very low-mass planets and on the longer term, their habitability. The capability of population synthesis to produce output for direct falsifiability with future observations, i.e., its predictive power  is a strength of the method. Besides that, such predictions are also useful to estimate the yield of future instruments and surveys.
		
\section{Global planet formation and evolution models}\label{sect:INTROglobalmodel}
We now come to the description of the physics of global models, which is the central subject of this paper. On the observational side, usually only the initial conditions (the protoplanetary disks) and the final outcomes of the planet formation process (the planets) are accessible to observations, even if  in future direct imaging and \textsc{Alma} might allow to observe planet formation as it happens for a limited number of cases \citep[e.g.,][]{quanzamara2013a}. With global models it is possible to bridge this gap at least on the theoretical side.  Many elements of modern planet formation theory that hold to this day were  first developed  by  \citet{safronov1969}. Other works that laid the foundations for the theoretical descriptions used here include \citet{shakurasunyaev1973,lynden-bellpringle1974,perricameron1974,mizunonakazawa1978,goldreichtremaine1979,hayashi1981,linpapaloizou1986a,bodenheimerpollack1986} and \citet{lissauer1993}.

Global numerical models of planet formation \citep[like][]{idalin2004a,idalin2008,idalin2008c,alibertmordasini2005,thommesmatsumura2008a,hellarynelson2012,mordasinialibert2012a} try to cover all major mechanisms that govern the formation and evolution process, and try to follow the formation process from its beginning to its end. The various mechanisms like accretion or migration must be treated in an interlinked way because they happen on similar timescales and feed back into each other.  Ideally, the models would start with a protoplanetary disk at a very early stage when the solids are in the form of micrometer-sized dust grains (or in principle even earlier with the collapse of the cloud), and yield as an output full-blown planetary systems with fully characterized planets at an age of several billions of years. In reality this is impossible, but it means that also the evolution of the planets over long timescales should be modeled, since in  most  cases, we observe planets a long time after they have  formed.  This applies both to the long-term evolution of the internal structure of a planet (its cooling and contraction) and to the secular evolution of the orbits due to gravitational interactions  {and tides}.

The fact that the models must follow the planetary systems during the entire formation process   is  the reason why global models are typically one (or at most two) dimensional, for example, in the description of the protoplanetary disk (assumed to be axisymmetric) or the internal structure of the planets (assumed to be spherically symmetric). More realistic 2D or 3D hydrodynamical simulation can at least currently not be used to simulate thousands of different initial conditions (protoplanetary disks) over their entire lifetime.

Most global models used to date in population synthesis calculations are  based on the  core accretion paradigm \citep{perricameron1974,mizunonakazawa1978}.  Core accretion states that first, solid cores form, some of which later accrete massive gaseous envelops to become giant planets (bottom-up process). The remaining cores collide to form both ice giants and terrestrial planets \citep[for an overview of this sequential picture of planet formation, see, e.g.,][]{papaloizouterquem2006,mordasiniklahr2010}. First statistical considerations based on the competing gravitational instability model where giant planets form directly from a gravitational instability in the protoplanetary gas disk \citep{cameron1978,boss1997} were recently made in \citet{jansonbonavita2011,jansonbonavita2012,forganrice2013} and \citet{galvagnimayer2013}. 

Global models address the different physical processes in a number of interlinked  sub-models. Figure \ref{fig:INTROschema}  {lists} the eleven sub-models of the combined formation (\citealt{alibertmordasini2005,alibertcarron2013}) and evolution model \citep[][]{mordasinialibert2012a, mordasinialibert2012}, on which we concentrate in the following. However, we also compare with other global models and review important ongoing and future work. Each sub-model is relatively simple, but the interaction of them leads to a considerable complexity. We next discuss the physics included in the different sub-models.  {They can be split in three classes: models for the protoplanetary disk, for one (proto)planet, and for the interactions (migration and N-body interaction).}

\subsection{Vertical structure of the protoplanetary disk}
A first sub-model calculates the structure and evolution of the gaseous protoplanetary disk \citep{lynden-bellpringle1974,papaloizouterquem1999,alibertmordasini2005}.  The gaseous disk model yields the ambient properties in which the planets form. The ambient pressure and temperature  serve as outer boundary conditions for the calculation of the structure of the gaseous envelope of the planets. The structure of the disk is also very important for the orbital migration of the protoplanets, since the direction and migration rate depends on the radial slopes of the temperature and the gas surface density (Sect. \ref{sect:INTROorbitalmig}). A good compromise for the numerical description of the disks  that are in reality very complex, 3D structures driven by (magneto-)hydrodynamical processes \citep[e.g.,][]{flockdzyurkevich2011} is provided by $\alpha$-viscosity models \citep{shakurasunyaev1973} that describe the disk as a rotating viscous fluid.  In the model of \citet{alibertmordasini2005}, the disk is described in the usual 1+1D approximation, i.e., the disk has a vertical and a radial structure, but is assumed to be axisymmetric. 

The vertical structure as a function of height $z$ above the disk midplane is obtained by solving the coupled equations of hydrostatic equilibrium, energy conservation, and energy transfer in the diffusion approximation for the radiative flux described by the equations \citep[e.g.,][]{papaloizouterquem1999}

 \begin{align}
\frac{1}{\rho}\frac{\partial P}{\partial z}&=-\Omega^{2}z &\frac{\partial F}{\partial z}&=\frac{9}{4} \rho \nu \Omega^{2} & F&=\frac{-16 \sigma T^{3}}{3 \kappa \rho}\frac{\partial T}{\partial z}.
\end{align}
The first equations is the hydrostatic equilibrium in a thin disk at a pressure $P$ and density $\rho$ for a Keplerian frequency $\Omega=\sqrt{G M_{\star}/r^{3}}$ where $G$ is the gravitational constant, $M_{\star}$ the mass of the star, and $r$ is the orbital distance. The second and third equation state that the energy liberated by viscous dissipation (characterized by a turbulent viscosity $\nu$) causes a flux of energy $F$ that is transported via radiative diffusion to be radiated away at the surface of the disk. The other quantities in the equation are the temperature $T$ and the opacity $\kappa$, while $\sigma$ is the Stefan-Boltzmann constant.

The turbulent viscosity is calculated using the classical $\alpha$ parametrization of  \citet{shakurasunyaev1973} as  {($c_{\rm s}$ is the sound speed)} $\nu=\alpha c_s^{2}/\Omega$  {in which all the complex physics about the angular transport processes in protoplanetary disks is hidden. The current understanding about these processes is actually still quite poor; for a recent review, see \citet{turnerfromang2014}}.  {In the context of the vertical disk structure, we note that the $\alpha$ parametrization of viscosity leads to viscous heating that is concentrated around the disk midplane. This is a consequence of  the vertically constant $\alpha$. Direct radiation-magnetohydrodynamic simulation where the turbulence is   {due to} the magneto-rotational instability (MRI, \citealt{balbushawley1991}) lead in contrast to a less concentrated heating, and thus to a different vertical temperature profile than in an $\alpha$ model \citep{flockfromang2013}.  Such a difference is particularly strong when the turbulence is primarily occurring above the midplane  {as it is the case in an MRI active layer above a midplane deadzone} \citep[e.g.,][]{dzyurkevichturner2013}. This (and other differences, see \citealt{turnerfromang2014}) show that the $\alpha$ parametrization of the turbulence might not be sufficient for a realistic description of protoplanetary disks.} 

Viscous dissipation is the dominant heating mechanism  in the inner parts of the disk \citep[e.g.,][]{chambers2009}. At larger orbital distances, the irradiation of the host star becomes dominant. This heating source can be incorporated into the surface boundary condition of the temperature  structure $T_{\rm s}$ as \citep{barriere-fouchetalibert2012}
\beq
T_{\rm s}^{4}=T_{\rm s,vis}^{4}+T_{\rm s,irr}^{4}
\eeq
where $T_{\rm s,vis}$ would be the temperature due to viscous heating only. The temperature due to the stellar irradiation is approximated as \citep{huesoguillot2005} 
\beq
T_{\rm s,irr}=T_{\star}\left[\frac{2}{3\pi}\left(\frac{R_{\star}}{r}\right)^{3}   +\frac{1}{2} \left(\frac{R_{\star}}{r}\right)^{2}   \left(\frac{H}{r}\right)   \left(\frac{d \ln H}{d \ln r}-1\right)  \right]^{1/4}
\eeq
where $T_{\star}$,  $R_{\star}$, and $H$ are the stellar temperature, radius, and the vertical pressure scale height, respectively.  \citet{barriere-fouchetalibert2012} set the flaring angle $d \ln H / d \ln r$ for simplicity to the equilibrium value of 9/7 \citep{chianggoldreich1997} which, however, means that the flaring angle and the possible effects of shadowing are not described in a self-consistent way. The solution of the vertical structure equations yield the disk midplane temperature and pressure, the vertical scale height $H$, and the vertically averaged viscosity.

\subsection{Radial structure of the protoplanetary (gas) disk}\label{sect:INTROradialdisk}
The evolution of gas surface density $\Sigma$ as a function of distance $r$ and time $t$ is described by the classical viscous evolution equation  of \citet{lynden-bellpringle1974} (first term on the right-hand side) supplemented by the effects of mass loss by photoevaporation  $\dot{\Sigma}_w(r)$ and accretion onto the planet $\dot{\Sigma}_{\rm planet}(r)$
\beq
\frac{\partial \Sigma}{\partial t}=\frac{1}{r}\frac{\partial}{\partial r}\left[3 r^{1/2} \frac{\partial}{\partial r}\left(r^{1/2}\nu \Sigma\right)\right]-\dot{\Sigma}_w(r)-\dot{\Sigma}_{\rm planet}(r).
\eeq

The mass loss due to photoevaporation \citep[for a recent review  {on disk dispersal mechanisms}, see][]{alexanderpascucci2013} has two origins \citep[][]{mordasinialibert2012}: first external photoevaporation due to far-ultraviolet (FUV) radiation coming from massive stars in the vicinity of the host star \citep[e.g.,][]{matsuyamajohnstone2003}. This drives a wind approximately outside of a gravitational radius $R_{g,{\rm I}}$ given as
\beq
R_{g,{\rm I}}=\frac{G M_*}{c_{s,{\rm I}}^2}
\eeq
where $c_{s,{\rm I}}$ is the speed of sound in the heated layer ($T_{\rm I}\approx1000$ K) of dissociated neutral hydrogen.  For a solar-like star, $R_{g,{\rm I}}$ is about 140 AU. The reduction of the surface density for a disk with total radius $r_{\rm max}$ is given as 
\beq
\dot{\Sigma}_{w,{\rm ext}}  =\left\{ \begin{array}{ll}
0 & \textrm{for}\  r < \beta_{\rm I} R_{g,{\rm I}} \\
\frac{\dot{M}_{\rm wind, ext}}{\pi (r_{\rm max}^2- \beta^2 R_{g,{\rm I}}^2)} & \textrm{otherwise,}
\end{array} \right.
\eeq
i.e., it occurs outside of an effective gravitational radius $\beta_{\rm I} R_{g,{\rm I}}$. The total rate $\dot{M}_{\rm wind, ext}$ is  {an input} parameter, and one of the Monte Carlo random variables in a population synthesis calculation. Its distribution function is chosen in a way that the resulting distribution of lifetimes of the synthetic disks is in agreement with the observed distribution of lifetimes of actual protoplanetary disks \citep{haischlada2001}. Physically, it depends on the number of and distance to massive stars in the vicinity of the host star \citep{adamshollenbach2004}. 

The second contribution to $\dot{\Sigma}_w(r)$ is due to internal photoevaporation driven by the EUV radiation of the host star itself. This ionizing radiation heats the surface layers to a temperature of $\sim$10$^{4}$ K which launches a wind with a velocity  $c_{s,{\rm II}}$ outside around $\beta_{\rm II}  R_{g,{\rm II}}$ which corresponds to approximately 7 AU for a solar-like star. The mass loss rate is 
\beq
\dot{\Sigma}_{w,{\rm int}}  =\left\{ \begin{array}{ll}
0 & \textrm{for}\  r < \beta_{\rm II}  R_{g,{\rm II}}\\
2 c_{s,{\rm II}} n_0(r) m_{\rm H} & \textrm{otherwise},
\end{array} \right.
\eeq
where $m_{\rm H}$ is the mass of  ionized hydrogen. The density of ions at the base of the wind $n_0(r)$ as a function of distance is approximately \citep{hollenbachjohnstone1994}.
\beq
n_{0}(r)=n_{0}(R_{\rm 14}) \left(\frac{r}{\beta_{\rm II}  R_{g,{\rm II}}}\right)^{-5/2}
\eeq
which means that most of the wind is originating close to the effective gravitational radius. The density  $n_{0}$ at a normalization radius $R_{\rm 14}$ is found from radiation-hydrodynamic simulations \citep{hollenbachjohnstone1994}.  {It increases with the square root of} the ionizing photon luminosity of the central star. 

The last term in the master equation for the evolution of $\Sigma$ is calculated by assuming that the gas accreted by a planet at a rate $\dot{M}_{XY}$
is removed from an annulus around it with a width equal to the planet's Hill sphere so that
\beq
\dot{\Sigma}_{\rm planet}=\frac{\dot{M}_{XY}}{2 \pi a R_{\rm H}}.
\eeq
The Hill sphere radius $R_{\rm H}$ of a planet of mass $M$ and semimajor axis $a$ around a star of mass $M_{\star}$ is given as $(M/3 M_{\star})^{1/3}a$. One finds that this term is important for the global evolution of the disk only for massive planets undergoing gas runaway accretion. 

A surface density at time $t=0$ must be specified as initial condition\footnote{It is clear that this is an artificial concept since disks are a gradually forming byproduct of star formation. Simple cloud collapse models \citep{shu1977} can be used to describe the initial formation of the disk by infall \citep{huesoguillot2005,dullemondapai2006}.}. Early population synthesis calculations \citep{idalin2004a,mordasinialibert2009} typically used an initial radial profile inspired by the minimum mass Solar nebula (MMSN) \citep{weidenschilling1977,hayashi1981} which is a power-law for $\Sigma$ falling as $r^{-3/2}$. Recent sub-millimeter observation of protoplanetary disks \citep[e.g.,][]{andrewswilner2010} rather indicate profiles  like  
\beq\label{eq:INTROsigmart0}
\Sigma(r,t=0)=\Sigma_0\left(\frac{r}{R_0}\right)^{-\gamma}\exp\left[-\left(\frac{r}{R_c}\right)^{2-\gamma}\right]
\eeq
where the mean values of the parameters $\gamma$ and $R_c$ are  approximately 0.9 and 30 AU, respectively \citep{andrewswilner2010}. Such profiles were used as initial conditions in more recent syntheses like \citet{mordasinialibert2012} and \citet{alibertcarron2013} even though the parameters were strictly speaking determined at distances outside of $\sim$20 AU, i.e., outside of the main planet formation region. The initial surface density $\Sigma_{0}$ at a normalization distance $R_{0}$ determines the initial total mass and is another Monte Carlo random variables in population syntheses.

\begin{figure}
\begin{center}  
{\includegraphics[width=0.95\columnwidth]{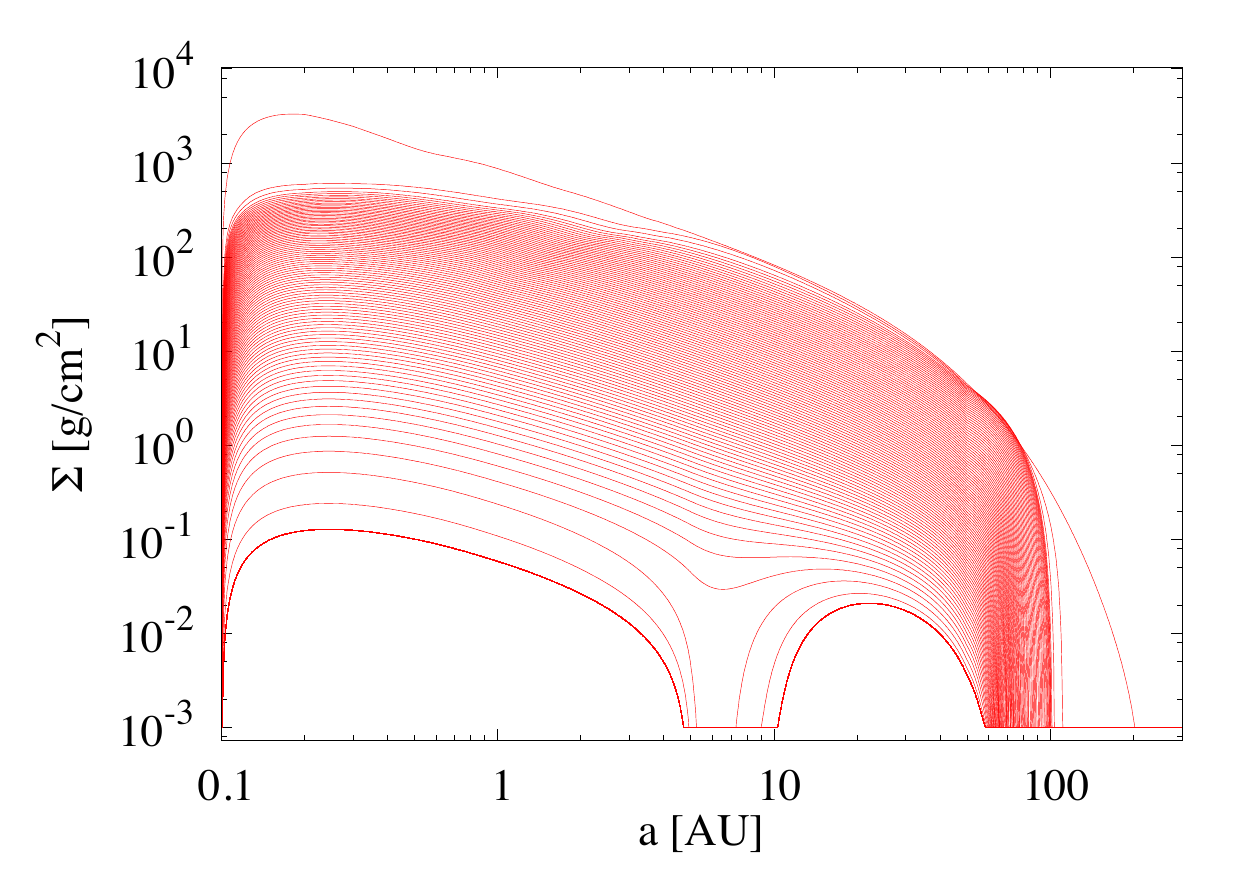}}
\caption{Example of the evolution of the gas surface density $\Sigma$ as a function of time and radius $a$. The uppermost line shows a state close to the beginning of the simulation and is similar to the initial condition given by Eq. \ref{eq:INTROsigmart0}.  Subsequent profiles are plotted 2$\times10^{4}$ years apart \citep[figure from][]{mordasinialibert2012}.} \label{fig:INTROsigma} 
\end{center}
\end{figure}

Figure \ref{fig:INTROsigma} shows a typical example of the evolution of the gas surface density under the action of viscosity and photoevaporation (but without an embedded planet).
The inner disk edge is fixed to 0.1 AU while the outer radius is free to spread or shrink. The initial mass of the disk is similar to the MMSN.

The global models of \citet{idalin2004a} and \citet{hellarynelson2012} use descriptions of the protoplanetary gas disk that are more  {parameterized. Typically, one assumes} a power-law dependency of the gas surface density as a function of radius all the time (i.e., not only as initial condition) and an exponential decrease on a uniform timescale $\tau_{\rm disk}$. For a power-law exponent $p_{\Sigma}$, the gas surface density is  then given as
\beq
\Sigma(r,t)=\Sigma_{0}\left(\frac{r}{R_{0}}\right)^{-p_{\Sigma}}e^{-t/\tau_{\rm disk}}.
\eeq
The radial temperature profile of the disk is described in an analogous way, that is as a power-law with a constant exponent $p_{\rm T}$. These exponents must be carefully chosen since the migration of low-mass planets particularly in non-isothermal parts of the disk depends in a sensitive way on the local values of $p_{\Sigma}$ and $p_{\rm T}$ \citep[][Sect. \ref{sect:INTROorbitalmig}]{lyrapaardekooper2010,kretkelin2012b,dittkristmordasini2014}.  In 1+1D $\alpha$ models, the exponents are functions of distance and time (they depend via the vertical structure on opacity transitions), which leads in particular to convergence zones that act as traps for migrating planets \citep[e.g.,][see also Sect. \ref{sect:INTROorbitalmig}]{paardekooperbaruteau2010}.

This does, however, not mean that 1+1D $\alpha$ models are already sufficient to catch all important effects found in 3D magnetohydrodynamical simulations.  {Regarding the general evolution and structure of the disk itself, an important assumption in the models presented here is the one of a radially constant $\alpha$.  The results obtained for the radial disk structure (like the surface density or temperature) depend significantly on the magnitude and assumed radial profile of $\alpha$ \citep[e.g.,][]{bellcassen1997,kretkelin2010}. It is clear that in actual protoplanetary disks, various processes occurring at different radii like the strength of the MRI, the radial and vertical extent of a possible low-turbulence deadzone \citep[e.g.,][]{dzyurkevichturner2013}, or the occurrence of layered accretion \citep{gammie1996} produce radial variations in the effective $\alpha$.  The use of one constant $\alpha$ must therefore be seen as a strong simplification.}

Regarding planet migration,  {changes in the effective $\alpha$ can significantly affect  planetary migration tracks by slowing or stopping the migration in planet traps \citep[e.g.,][]{matsumurapudritz2007}. In the context of global planet formation models, this was studied in \citet{idalin2008} and \citet{hasegawapudritz2013}. Furthermore,}  \citet{uribeklahr2011} find additional torques in 3D magnetohydrodynamical simulations where the turbulence is given by the magneto-rotational instability.  They are directly related to the presence of the magnetic fields so that they cannot be described in a 1+1D $\alpha$ model.

The output of the disk structure model is used in several other sub-models. The ``atmosphere'' model, for example, needs the midplane pressure and temperature in the nebula to calculate the boundary conditions for the planet's interior. The ``migration'' models needs, as mentioned, the surface density and temperature gradients, the surface density itself, the vertical scale height $H$, and the viscosity to calculate the planetary migration rate. 

\subsection{Disk of solids (planetesimals, fragments)} 
A third sub-model describes the structure and evolution of the disk of  small solid bodies (planetesimals) from which the bigger protoplanets grow. The division in ``small'' and ``big'' bodies is made since runaway growth leads to a bimodal size distribution \citep{weidenschillingspaute1997}.  The model of the disk of solids yields as a function of time and orbital distance the typical size $s_{\rm solid}$ (or size distribution) of the bodies, their dynamical state (random velocities or, equivalently, inclinations  and eccentricities) and their surface density $\Sigma_{\rm solid}$. These quantities control the core accretion rate  of a forming protoplanet. 

In a protoplanetary disk, the solids are initially in the form of tiny dust grains that grow in time to form kilometer-sized planetesimals \citep[by a process that is currently debated, e.g.,][]{cuzzihogan2008}.  Destructive collisions can again reduce the size of the bodies. Additionally, solid bodies drift radially through the disk at a rate that depends among others things on the size of the bodies.  The dynamical state of the small bodies is influenced by several non-linear processes like dynamical friction, viscous stirring, or damping by gas drag \citep[e.g.,][]{idamakino1993,ormeldullemond2010}. The interaction of all this processes make that the evolution of the disk of solids is in reality a very complex problem. Some aspects (in particular the viscous stirring by the protoplanets) were recently included in global models by \citet{fortieralibert2013}. In earlier population syntheses, \citet{mordasinialibert2009}  {used} a very simple model  {for} the solids where the disk of planetesimals only changes due to the accretion and ejection of planetesimals by the protoplanet. The size of the planetesimals is uniform in time and space (usually 100 km).  It is furthermore assumed that the accretion and ejection (for massive protoplanets) homogeneously reduces the surface density of planetesimals in the feeding zone of the planet, which is  {taken} to have a radial width equal to $W_{\rm feed}=B_{\rm L} R_{\rm H}$ where $B_{\rm L}\approx 4-5$ \citep{lissauer1993}.   {This ignores the potential formation of a gap in the disk of planetesimals \citep{shiraishiida2008}.} The decrease of the planetesimal surface density is then simply \citep{thommesduncan2003}
\beq
\frac{d \Sigma_{\rm solid}}{dt} = - \frac{(3 M_\star)^{1/3}}{6\pi a^2 B_{\rm L} M^{1/3}}\frac{d M_{\rm Z,tot}}{dt}
\eeq
where $d M_{\rm Z,tot}/dt$ denotes the total change of surface density of planetesimals due to the presence of the protoplanet, i.e., the sum of the rate at which the planet accretes and ejects planetesimals. For the initial condition it is assumed that the surface density of planetesimals is proportional to the initial surface density of the gas times the dust-to-gas ratio $f_{\rm D/G}$ as 
\beq
\Sigma_{\rm solid}(r,t=0)=f_{\rm D/G} f_{\rm R/I} \Sigma(r,t=0)
\eeq

\begin{figure}
\begin{center}  
{\includegraphics[width=0.95 \columnwidth]{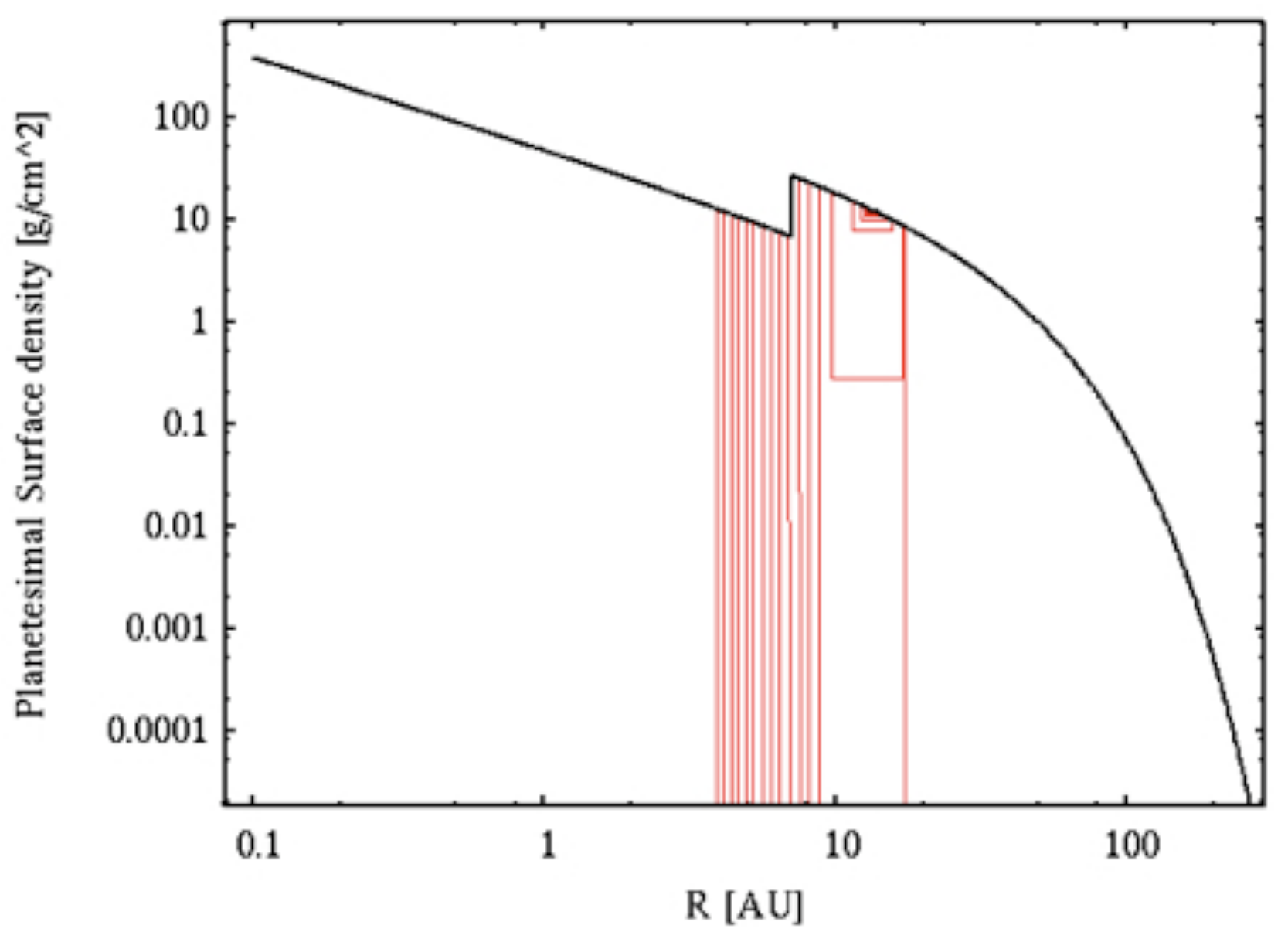}}
\caption{Surface density of planetesimals.  The black line shows the initial planetesimal surface density in the entire disk. The  increase at the iceline is visible. The red lines show the reduction of $\Sigma_{\rm solid}$ due to the presence of an  {accreting and} migrating planet.  } \label{fig:INTROsigmasolid} 
\end{center}
\end{figure}

The dust-to-gas ratio is another Monte Carlo random variable in population synthesis calculation, and set under the simplifying assumption that it is directly given  by the stellar [Fe/H] as $f_{\rm D/G}=f_{\rm D/G,\odot} \times 10^{\rm [Fe/H]}$ where $f_{\rm D/G,\odot}$ is the solar metal fraction. The factor $f_{\rm R/I}$  {represents} the increase of the surface density due to the condensation of water ice outside of the iceline, and set to \citep{hayashi1981}
\beq
f_{\rm R/I} =\left\{ \begin{array}{ll}
1 & \textrm{for}\  T_{\rm mid}(t=0) < 170 K  \ \ (a>a_{\rm ice})\\
1/4 & \textrm{otherwise}\ \ \ \ \ \ \ \ \ \ \ \ \ \ \ \ \ \ \ \ \, (a<a_{\rm ice})
\end{array} \right. 
\eeq
which means that the initial midplane temperature profile determines the location of the iceline $a_{\rm ice}$.  {More recent works on the composition of the solar nebula \citep[like][]{lodders2003} indicate that the change of the surface density at the iceline is probably rather about a factor of 2 than 4 as assumed  by  \citet{hayashi1981}.}  Figure \ref{fig:INTROsigmasolid} shows the initial surface density and its evolution due to a growing protoplanet.  The initial profile simply traces the initial profile of the gas (Eq. \ref{eq:INTROsigmart0}) except for the sudden increase at the iceline. The planet initially forms at about 15 AU, and then migrates inwards depleting the planetesimal disk down to a distance of about 5 AU. Due to migration, the expansion of the feeding zone occurs mainly towards the interior part of the disk. It occurs on a shorter timescale than for in situ formation, which can reduce the overall formation timescale of a giant planet \citep{alibertmordasini2004}.

Other global planet formation models \citep{idalin2004a,hellarynelson2012} use  similar  prescriptions for the initial surface density distribution. The model of \citet{idalin2008} is more complex because it takes into account that  {a} magnetohydrodynamically inactive  deadzone in the protoplanetary disk can lead to a local density maximum of planetesimals in the vicinity of the iceline. But it is clear that all global models treat  {to date} the processes occurring during the early condensation of the solids and the subsequent  growth phase from dust to planetesimals in a simplistic way (or not at all). 

{A complete discussion of the key physical processes associated with the formation and growth of planetesimals is beyond the scope of this work (see \citealt{johansenblum2014} for a recent review).  {There is currently no generally accepted complete theory for the  growth from $\mu$m-sized dust to sizes where gravitational focusing sets in, allowing efficient runaway growth (of order 1-1000 km, \citealt{ormelokuzumi2013}).} An incomplete list of important issues that must be addressed includes: (i) the basic mode of growth which is two-body coagulation or self-gravitational instability \citep{birnstieldullemond2010,johansenoishi2007}; these are not mutually exclusive processes \citep{johansenblum2014}, (ii) the mode of concentration of small bodies in the turbulent disk on large \citep{johansenklahr2006} or small scales \citep{cuzzihogan2008}, (iii) the strong dependency of planetesimal formation rates on the dust-to-gas ratio \citep[e.g.,][]{johansenyoudin2009,cuzzihogan2010} so that planetesimals may form preferentially at certain disk locations rather than uniformly (as assumed in the global models), (iv) the material properties of dust grains and planetesimals and their behavior during mutual collisions which are very poorly constrained \citep{guettlerblum2010,leinhardtstewart2012}, (v)  the possibility that planetesimal formation may be inefficient or happen over an extended period of time (several Myrs) based on observations of chondrite parent bodies \citep{bizzarrobaker2004}, in contrast to the assumptions in the global models,  (vi) the possibility that much of the solid mass existed in mm-to-m size particles that drift radially, significantly  modifying the surface density profile, the solid-to-gas and rock-to-ice ratio over time \citep[e.g.,][]{cuzzizahnle2004,cieslacuzzi2006}, also in contrast to the simplifications in the global models,  (vii)  the mechanism that  planetesimal-planetesimal collisions may rapidly convert most of the solid mass into smaller objects  \cite[e.g.,][]{inabawetherill2003,chambers2008} so that at least three types of bodies must be included in the solid disk model (embryos, planetesimals, and fragments, see \citealt{ormelkobayashi2012}), and finally that the timescale and even existence of runaway growth to form  protoplanets depends sensitively on the turbulence level in the disk due to eccentricity excitation by turbulent density fluctuations \cite[e.g.,][]{nelson2005,idaguillot2008,ormelokuzumi2013}.} 

 {Once a clearer and more unified picture of planetesimal formation  {arises}, simplified version of it will again be included in the global models to understand the observational consequences. Some steps towards this can be  found in the aforementioned works and, e.g., \citet[][]{kornetwolf2007,carter-bondobrien2010,birnstielklahr2012} or \citet{chambers2014}. }

\subsection{Planetary core accretion rate}\label{sect:INTROsolidaccretionrate}
The solid core of a protoplanet grows by the accretion of  background planetesimals and by the collision with other protoplanets in the case that several planets form concurrently. The simplest way to describe the collisional growth of protoplanet due to the accretion of planetesimals is a Safronov-type rate equation \citep{safronov1969} for the  accretion rate of the core of mass $\mz$
\beq\label{eq:safronov}
\frac{d \mz}{dt}=\Omega\Sigma_{\rm solid} R_{\rm capture}^2 F_{\rm G}(e,i).
\eeq
The capture radius $R_{\rm capture}$ for planetesimals is larger than the core radius due to the presence of the gaseous envelope. It is calculated in the ``infall'' model, while the $\Sigma_{\rm solid}$ is an output of the model discussed in the last section. The gravitational focussing factor  $F_{\rm G}(e,i)$, that  takes into account 3-body effects (planet, planetesimal, and star), is given  {for example} by \citet{greenzweiglissauer1992}. Its magnitude depends on the dynamical state of the planetesimal swarm.  {This} is the key quantity  determining  $d \mz/dt$ since it gives raise to different growth regimes like runaway, oligarchic or orderly growth, see, e.g., \citet{rafikov2003d}. In the population syntheses of Mordasini et al. (\citeyear{mordasinialibert2009})-(\citeyear{mordasinialibert2012})  the original expressions of \citet{pollackhubickyj1996} are used to estimate the eccentricities and inclinations. These equations predict low random velocities, allowing rapid growth, which is probably difficult to achieve for 100 km planetesimals  {in reality} \citep[e.g.,][]{thommesduncan2003}. An updated description with a  more detailed description of the interactions between protoplanet and planetesimals  {can be found}  in \citet{fortieralibert2013}, but it is likely that the prescriptions for the core accretion rate will be further  {significantly} modified  in view of new result of specialized models:  {as understood recently \citep[e.g.,][]{ormelklahr2010,lambrechtsjohansen2012,morbidellinesvorny2012,chambers2014} the accretion rate of decimeters-sized pebbles instead of km-sized planetesimals can be very high, especially also at larger semimajor axes. Such high rates are necessary to form a $\sim$10 $\mearth$ core during the presence of the nebula in a MMSN disk which is otherwise far from simple to achieve, see, e.g., \citet{ormelokuzumi2013} or \citet{chambers2014}. The high accretion rates are due to a strongly increased  capture cross section for small particles due to drag with the disk gas during the encounter with the protoplanetary core. Understanding the global predictions of pebble accretion when coupled to global planet formation models and population syntheses  is an important task for future studies.}

The formation model of \citet{idalin2004a} also uses a Safronov-type rate equation but describes the dynamical state of the swarm  based on more recent results for the oligarchic growth regime \citep{idamakino1993,ormeldullemond2010}. As they assume smaller planetesimals (km-sized instead of 100 km), the resulting  core growth timescales are nevertheless relatively similar \citep{mordasinialibert2009a}. The global model of \citet{hellarynelson2012} is much more detailed  because the solid accretion rate is found by direct N-body integration of a few ten protoplanets plus a several thousand planetesimals. The downside of this is that the associated long computation time makes the calculation of statistically sufficiently large population of synthetic planets difficult, at least for the moment.  

\subsection{Internal structure of the planetary gas envelope}\label{sect:INTROinternalstruct}
The sub-models discussed up to this point describe the protoplanetary disk and the growth of the core. The ``envelope'' model (together with the internal structure model of the core) deals with the protoplanet itself,  calculating the internal 1D (spherically symmetric) radial structure of the gaseous envelope (H/He) of the planet. During the formation phase, the model in particular yields the gas accretion rate $\dot{M}_{\rm XY}$. Low-mass cores can only bind tenuous atmospheres, while cores more massive than roughly 10 Earth masses can trigger rapid runaway gas accretion, so that a giant planet forms. After the formation phase the long-term evolution, i.e., the contraction and cooling at constant mass is calculated, which yields the radius and intrinsic luminosity of a planet  {(recent reviews on planetary internal structures and the temporal evolution can be found in \citealt{baraffechabrier2014} and \citealt{chabrierjohansen2014})}. The internal structure is found by integrating the standard equations for planetary interiors \citep[e.g.,][]{bodenheimerpollack1986} which are the equations of mass conservation, hydrostatic equilibrium, energy conservation and energy transfer:

\begin{alignat}{2}
\frac{\partial m}{\partial r}&=4 \pi r^{2} \rho    &\quad  \quad \frac{\partial P}{\partial r}&=-\frac{G m}{r^{2}}\rho    \\
\frac{\partial l}{\partial r}&=4 \pi r^{2} \rho\left(\varepsilon -P \frac{\partial V}{\partial t} -\frac{\partial u}{\partial t}\right)             & \frac{ \partial T}{\partial r}&=\frac{T}{P}\frac{\partial P}{\partial r}\nabla(T,P)          
\end{alignat}

\begin{figure*}
\begin{center}  
{\includegraphics[width=0.99\textwidth]{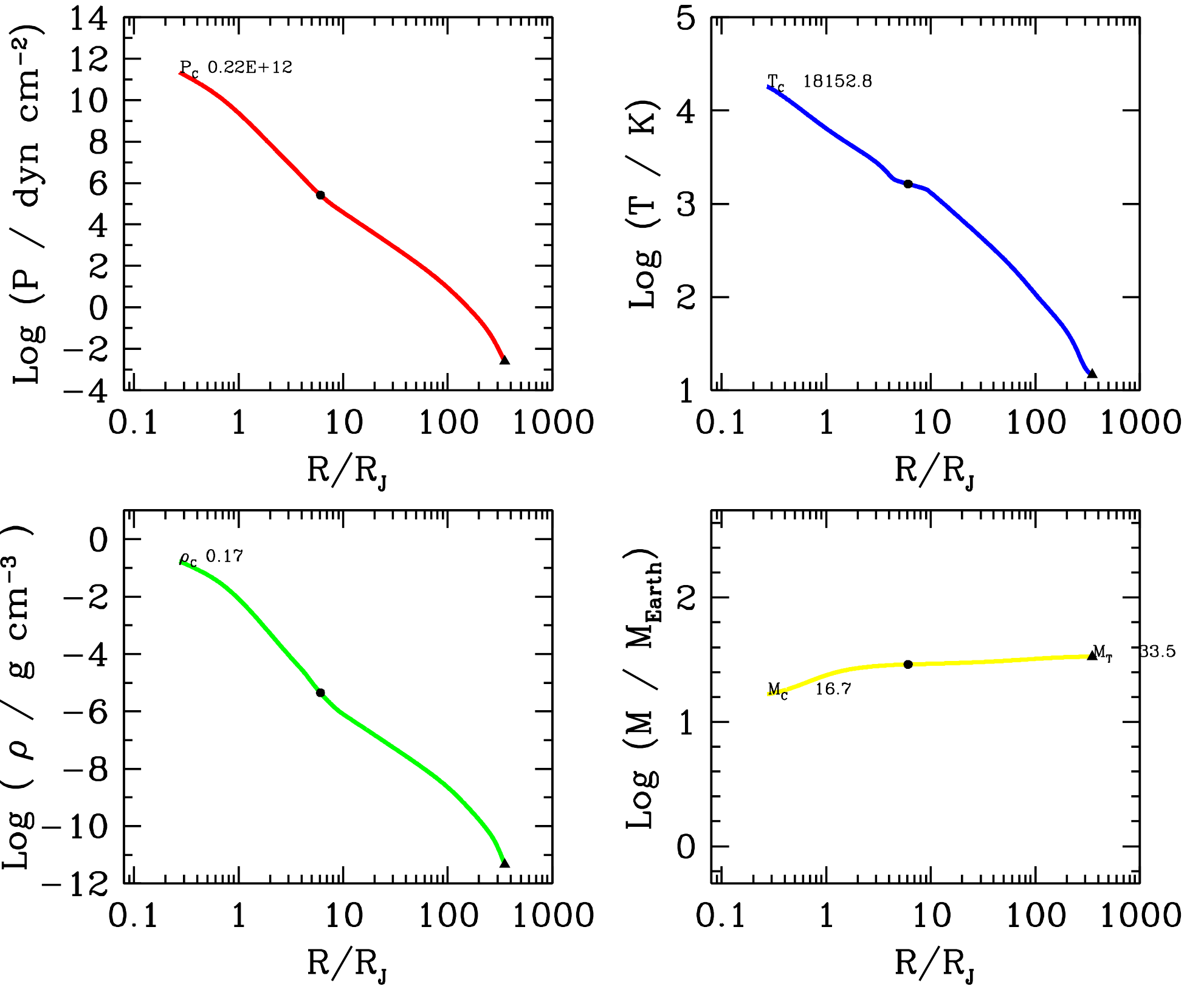}}
\caption{Radial envelope structure inside a growing giant planet at ``crossover''.   The four panels show the pressure, temperature, density, and mass as function of radius. The values at the core-envelope boundary are also given. The black dot on the lines indicates the position of the capture radius for 100 km planetesimals.   } \label{fig:INTROstructcross} 
\end{center}
\end{figure*}

In these equations, $r$ is the radius as measured from the planetary center, $m$ the gas mass inside $r$, $\rho, P, T, u$ the gas density, pressure, temperature, and specific internal energy, $V$ the specific volume  $1/\rho$,  {and} $t$ the time.  {T}he gradient $\nabla=d \ln T / d \ln P $ is either the radiative or the adiabatic gradient, whichever is shallower. The adiabatic gradient is directly given by the equation of state, while the radiative gradient is 
\beq\label{eq:radgrad}
\nabla_{\rm rad}=\frac{3}{64 \pi \sigma G}\frac{\kappa l P}{T^{4} m }.
\eeq
In this equation, $\sigma$ is the Stefan-Boltzmann constant while $\kappa$ is the Rosseland mean opacity. In the outer layers of the envelope, it is dominated by the opacity due to small grains suspended in the gas  {\citep[][Sect. \ref{sect:grainopacity}]{podolak2003,movshovitzbodenheimer2010,cuzziestrada2014}}.
Th {ese} are the same equations as for stellar interiors with the difference that the luminosity is not due to nuclear  burning of hydrogen but due to the contraction and cooling of the gas, and the energy deposition by impacting planetesimals that is found with the ``infall'' sub-model. The burning of deuterium in sufficiently massive objects  {can} also  {be} included in the energy source term  $\varepsilon$ \citep[][]{bodenheimerdangelo2013,mollieremordasini2012}. Note that the models of \citet{mordasinialibert2012a} and \citet{alibertcarron2013} simplify the energy equation by assuming a luminosity that is radially constant.

Figure \ref{fig:INTROstructcross} shows an example of the radial envelope structure inside  a growing giant planet that forms in situ at 5.2 AU, with initial conditions similar to the classical J1 simulations of \citet{pollackhubickyj1996}. The left end of the lines corresponds to the core-envelope interface at $R_{\rm core}$, while the outer radius approximately corresponds to the Hill sphere radius. The radial structure is shown at the ``crossover'' point which is the moment when the core and envelope mass are  equal. One sees that the pressure and density increase by many orders of magnitude across the envelope. At the outer edge, the density is of order $10^{-11}$ g/cm$^{3}$, which is typical for the outer parts of the solar nebula, while at the core-envelope interface, the density approaches an almost fluid like value of 0.17 g/cm$^{3}$. Also the temperature at this position is already quite  {high} with $T_{\rm c}\approx 18\ 000$ K. The plot also indicates the position of the protoplanet's capture radius for 100 km planetesimals. It is about ten times larger than the core radius, leading to a core accretion rate that is about a factor 100 larger compared to the case of an envelope free core (Eq. \ref{eq:safronov}).

At a small core mass, the gas accretion rate found by solving the internal structure equations $\dot{M}_{\rm KH}$ is small, because the Kelvin-Helmholtz timescale is long \citep{ikomanakazawa2000}. Once the crossover mass is reached which corresponds to the critical core mass in the early strictly static calculations \citep[like][]{perricameron1974,mizunonakazawa1978}, $\dot{M}_{\rm KH}$ starts to increase rapidly, and becomes at some moment ( {typically} when the total mass of the planet is of order 100 $\mearth)$ larger than the maximal rate $\dot{M}_{\rm XY, max}$ at which the protoplanetary disk can supply gas to the planet. This means that the envelope of the planet now contracts so quickly that (at least formally) an empty shell between the planet's envelope and the background nebula develops.  The planet therefore detaches from the nebula and contracts rapidly, becoming much more centrally condensed. The gas now freely falls from the Hill sphere onto the planet, where it is accreted through an accretion shock\footnote{This applies to the strictly 1D picture. Hydrodynamic simulation in 3D show a basically similar picture, with the difference that at the moment of rapid contraction, the spherical symmetry is lost, as a circumplanetary disk forms \citep{ayliffebate2012}. This can not be captured in 1D models.}.  The planetary gas accretion rate is thus
\beq 
\dot{M}_{\rm XY}=\min\left(\dot{M}_{\rm KH}, \dot{M}_{\rm XY, max} \right).
\eeq
The disk-limited gas accretion rate $\dot{M}_{\rm XY, max}$ is controlled by a number of processes, namely, the local availability of gas, the  {gas} flux through the disk that decreases itself in time, and the effect of gap formation \citep{lubowseibert1999,dangelodurisen2010}. Initially, the planet accretes rapidly the gas in its vicinity, so that the gas accretion rate can be approximate by a Bondi-type rate \citep{dangelolubow2008} 
\beq
\dot{M}_{\rm XY,max,R}\approx\frac{\Sigma}{H}\Omega R^3_{\rm gc} 
\eeq
where $R_{\rm gc}$ is the capture radius for gas which is approximately the smaller of (a fraction of) the Hill and the Bondi radius $G M/c_{\rm s}^{2}$ where $M$ is the planet's mass and $c_{\rm s}$ the sound speed in the background nebula. After the local reservoir has been exhausted, the global flux of gas in the disk towards the planet starts to limit the gas accretion. This rate is now given by \citep[e.g.,][]{mordasinialibert2012a}
\beq
\dot{M}_{\rm XY,max,F} =k_{\rm Lub} \left(3 \pi  \nu \Sigma + 6\pi r \frac{\partial \nu \Sigma}{\partial r}\right)
\eeq
where $k_{\rm Lub}$ is a factor that describes the fraction of the gas streaming viscously  {through the disk} towards the planet that is eventually accreted onto it \citep{lubowseibert1999,lubowdangelo2006}. Note that there is considerable uncertainty about the efficiency of gas accretion in the disk-limited regime \citep{benzida2013}. This directly influences the predictions for the upper end of the planetary mass function, but also for type II migration and the resulting formation of Hot Jupiters. 

Other global models of planet formation describe the accretion of gas  in the phase when it is limited by the envelope's contraction in a   {parameterized} way without solving the internal planet structure \citep{idalin2004a,thommesmatsumura2008a,hellarynelson2012}. Instead, a parametrization of the Kelvin-Helmholtz timescale $\tau_{\rm KH}$ is used, and the gas accretion rate is written in an equation of the form
\beq
\dot{M}_{\rm XY}=\frac{M}{\tau_{\rm KH}}
\eeq
where the Kelvin-Helmholtz timescale is itself a function of the planet mass $M$ and the opacity $\kappa$ \citep[][]{ikomanakazawa2000}, typically approximated as $ \tau_{\rm KH}=k M^{-p}  \kappa^{-q}$. A potential limitation of such an approach is the  dependency on the parameters $k$, $p$, and $q$ that are not well constrained  \citep[$p\approx1.9$-3.5,][]{miguelbrunini2008b}. This  {can have} directly visible consequences in population syntheses, for example in the predicted planetary mass function \citep{miguelbrunini2008b}.   {When} no internal structures are calculated, no direct predictions for transit or direct imaging searches can be made, especially at early times when the formation still directly influences these quantities. On the other hand, the computational cost is much reduced. It should be noted that also the solution of  the 1D  structure equations is (currently)  not completely free from prespecified parameters. This is due to the fact that in global models, the opacity $\kappa$ in Eq. \ref{eq:radgrad} is typically not found ab initio as in specialized models, but approximated as the ISM opacity multiplied with some prespecified reduction factor (cf. Sect. \ref{sect:grainopacity}).   {These specialized models \citep[][]{podolak2003,movshovitzpodolak2008,movshovitzbodenheimer2010} find the grain opacity self-consistently by solving the Smoluchowski equation in each atmospheric layer including the effects of grain growth, settling, and vaporization. This is computationally time consuming, therefore \citet{mordasiniklahr2014} have recently tried to parametrize these results  for use in population synthesis models by deriving the reduction factor of the ISM opacity that leads to gas accretion timescales that agree with the results of the specialized models. The problem with this approach is that one global reduction factor cannot capture the dependency of the mechanisms governing the grain dynamics on planetary properties like the core and envelope mass. } 

The ``envelope'', ``infall'', and ``accretion rate'' sub-models of the global formation model shown in Fig. \ref{fig:INTROschema} are very similar to classical 1D giant planet formation codes like \citet{bodenheimerpollack1986,pollackhubickyj1996} or \citet{lissauerhubickyj2009} which are the conceptual origin of this model  {(see \citealt{helledbodenheimer2013} for a recent review on giant planet formation)}. Despite this origin, most planets that form in a population synthesis are in fact low-mass planets that do not trigger gas runaway accretion. These planets can, however, still be modeled with the restriction that primordial  H/He envelopes are the only type of envelope that can currently be considered. 

\begin{figure}
\begin{center}  
\includegraphics[width=\columnwidth]{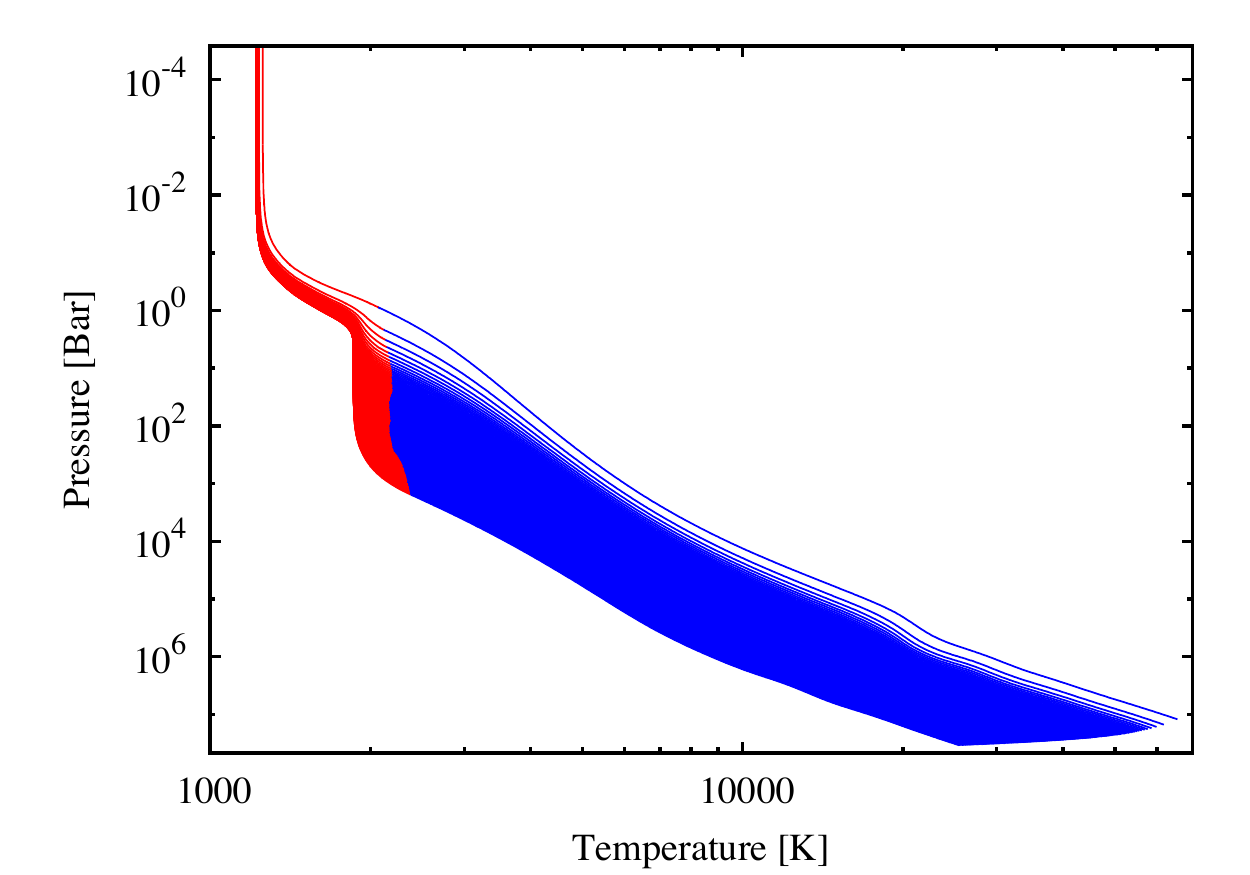}
\caption{Evolution of the interior and atmosphere over gigayears for a close-in giant planet. The uppermost line shows the pressure-temperature profile shortly after the end of formation. The lowermost line is the state after 5 Gyrs. The blue part of the lines corresponds to convective regions, while red indicates radiative zones. The mass of the planet is approximately one Jovian mass while the semimajor axis is 0.04 AU.} \label{fig:INTROptjupihot} 
\end{center}
\end{figure}

The extension into an evolutionary model \citep[][]{mordasinialibert2012a}  {can bring a} global model close to classical models of (giant) planet long-term evolution (cooling and contraction) like \citet{burrowsmarley1997} or \citet{baraffechabrier2003}. The set of  equations for the internal structure during the evolution after the dissipation of the protoplanetary nebula remains the same as during formation, but they are solved with different outer boundary conditions as described in the ``atmosphere'' section below. 

Figure \ref{fig:INTROptjupihot} shows the long-term evolution of the internal structure of a close-in hot Jupiter planet represented by its pressure-temperature profile. The figure shows a temporal series of profiles, including both the atmosphere and the complete interior. The upper end of the lines thus corresponds to the outer radius of the planet, while the lower  {is} the core-envelope boundary. The uppermost line shows the state shortly after the end of formation when the final mass has been reached, while the lowest line corresponds to the state after 5 Gyrs. The gradual cooling of the interior is obvious. The atmospheric model is the  {semi-gray} solution of \citet{guillot2010a} discussed below. The formation of a deep radiative zone that is characteristic for strongly irradiated giant planets is  visible  \citep{guillotshowman2002}. In contrast to the interior, the temperature at the outer radius is nearly constant since it is dominated by stellar irradiation, which was assumed to be constant (the temporal evolution of the host star and its luminosity is neglected).

\subsection{Atmosphere of the planet}\label{sect:INTROatmopla}
This sub-model provides the structure of the outer part of the (proto)planet and therefore also the boundary conditions necessary to solve the internal structure equations. These boundary conditions depend on the stage of formation or evolution a planet is in. Three major phases can be distinguished \citep[e.g.,][]{bodenheimerhubickyj2000}: 

During the first attached (or nebular) phase which applies to low-mass, sub-critical planets embedded in the protoplanetary nebulae, the envelope bound to the protoplanet smoothly transitions into the unbound, background conditions in the nebula. The structure extends out to a radius $R$ that is  {approximately} the smaller of the Hill or Bondi radius \citep{lissauerhubickyj2009}. No atmosphere in the classical sense exists\footnote{Sometimes, the outer radiative zone is called the ``atmosphere'' in this phase \citep[e.g.,][]{movshovitzpodolak2008}.}, and the boundary conditions are 

\begin{align}
T^{4}&=T_{\rm neb}^{4}+T_{\rm int}^{4} &P&=P_{\rm neb}\\
\tau&={\rm max}\left(\rho_{\rm neb}\kappa_{\rm neb}R, 2/3 \right)  &   T_{\rm int}^{4}&=\frac{3 \tau L_{\rm int}}{8 \pi \sigma R^2}. 
\end{align}
The pressure is  equal to the background midplane pressure in the nebula (provided by the disk model) $P_{\rm neb}$. The surface temperature contains  the contributions from the nebula midplane temperature  $T_{\rm neb}$ and from the intrinsic luminosity of the planet  $L_{\rm int}$. The effect of the optical depth $\tau$ of the background nebula is taken into account with the approximation of \citet{papaloizounelson2005}. 

The second detached (or transitional) phase applies to massive cores ($M_{\rm core}\gtrsim10\mearth$, total mass $M\gtrsim100\mearth$) that have triggered runaway gas accretion, i.e., where the gas accretion rate due to the contraction of the envelope $\dot{M}_{\rm KH}$ would be larger than the maximal rate at which the nebula can supply gas to the planet $\dot{M}_{\rm XY,max}$, as described above. The planet has now a free outer radius that collapses rapidly from initially the Hill sphere radius to about 2 to 3 $\rj$ (for cold accretion  {, i.e., low entropies, \citealt{marleyfortney2007}}). The gas falls approximately at free fall velocity $v_{\rm ff}$ from the Hill sphere onto the planet where it shocks. The boundary conditions are then \citep{bodenheimerhubickyj2000}
\begin{align}
  \mdotxy&=\dot{M}_{\rm XY,max}&   v_{\rm ff}^{2}&=2 G M \left(\frac{1}{R}-\frac{1}{R_{\rm H}}\right)\\
      P&=P_{\rm neb}+\frac{\mdotxy}{4 \pi R^{2}} v_{\rm ff}+\frac{2 g}{3 \kappa} & \tau&={\rm max}\left(\rho_{\rm neb}\kappa_{\rm neb}R, 2/3\right) \label{eq:pdetached} \\
    T_{\rm int}^{4}&=\frac{3 \tau L_{\rm int}}{8 \pi \sigma R^2} & T^{4}&=(1-A) T_{\rm neb}^{4}+T_{\rm int}^{4} \label{eq:tdetached}
\end{align} 
The outer pressure therefore contains the contributions from the background nebula, the accretion shock, and from the standard Eddington expression for the photospheric pressure due to the material residing above the $\tau=2/3$ surface. The temperature contains the contributions from the nebula and the intrinsic luminosity, where $A$ is the albedo of the planet.  {With} these boundary conditions  {one describes the accreting planet} as a  {scaled-down} version of a stellar hydrostatic core  {undergoing} spherical accretion as in  {the classical work of} \citet{stahlershu1980}.  {Hydrodynamic simulations indicate that in reality,} the actual  {infall} geometry and  {thus} the boundary conditions are more complicated  {than in this 1D picture},  {making} 3D radiation-hydrodynamic calculations   {necessary} \citep[e.g.,][]{klahrkley2006,ayliffebate2009}.

The third evolutionary (or isolated) phase  {starts} after the protoplanetary  {disk} has  {disappeared}. The planet  {now} evolves at constant mass ( {if we neglect} further accretion or mass loss through envelope evaporation for close-in planets, see Sect. \ref{sect:INTROescape}). \citet{mordasinialibert2012} model this phase with a simple gray atmosphere so that \citep[e.g.,][]{guillot2005}
\begin{align}
    P&=\frac{2 g}{3 \kappa} & T_{\rm int}^{4}&=\frac{ L_{\rm int}}{4 \pi \sigma R^2}\\
  T_{\rm equi}&=280\,{\rm K}  \left(\frac{a}{1 {\rm AU}}\right)^{-\frac{1}{2}}\left(\frac{\mstar}{\msun}\right) &  T^{4}&=(1-A) T_{\rm equi}^{4}+T_{\rm int}^{4}.\label{eq:tequievo}
\end{align}
The equilibrium temperature due to stellar irradiation is calculated assuming that the star with mass $\mstar$ is on the main sequence where the luminosity approximately scales as $\mstar^{4}$. Other evolutionary models like \citet{burrowsmarley1997,baraffechabrier2003,baraffechabrier2008a}  in contrast couple the interior calculation to full non-gray atmospheres.  

For a giant planet at a few AU where the irradiation flux from the host star is rather low, the gray atmosphere and the full non-gray atmospheres lead to similar cooling curves \citep{bodenheimerhubickyj2000,mordasinialibert2012}. For Hot Jupiter planets and in general all strongly irradiated planets, gray atmospheres lead, however, to too low temperatures deep in the atmosphere, so that the cooling timescale is underestimated \citep{guillotshowman2002}. A better atmospheric model than the gray atmosphere is  the  {semi-gray} approximation of \citet{guillot2010a}. This model provides the temperature as a function of optical depth in an atmosphere that transports both an intrinsic heat flux and receives an outer irradiation flux. The model is parametrized by two mean opacities, one in the visual and one in the infrared. The mean temperature as a function of IR optical depth $\tau$ is then  
\begin{dmath}
T^{4}=\frac{3 \tint^{4}}{4}\left(\frac{2}{3}+\tau\right)+\frac{3 T_{\rm equi}^{4}}{4}\left(\frac{2}{3}+\frac{2}{3\gamma}\left[1+\left(\frac{\gamma \tau}{2}-1 \right)e^{-\gamma \tau} \right]  \\  +\frac{2 \gamma}{3}\left(1-\frac{\tau^{2}}{2}\right)E_{2}(\gamma\tau)\right)
\end{dmath}
where $\gamma$ denotes the ratio of the mean opacity in the visual $\kappa_{\rm v}$ to the  mean opacity in the thermal infrared  $\kappa_{\rm th}$, while $E_{2}$ is an exponential integral. This expression provides a fair approximation of detailed irradiated atmosphere models \citep[e.g.,][]{showmanfortney2009} and therefore also gives more realistic cooling curves. This is important for the comparison of the radii of synthetic and actual transiting exoplanets. A better atmospheric structure also makes it  possible to calculate  the transit radius that is bigger than the $\tau$$\approx$1 radius to the grazing observational geometry more accurately \citep[e.g.,][]{hansen2008}.

\begin{figure}
\begin{center}  
{\includegraphics[width=0.99\columnwidth]{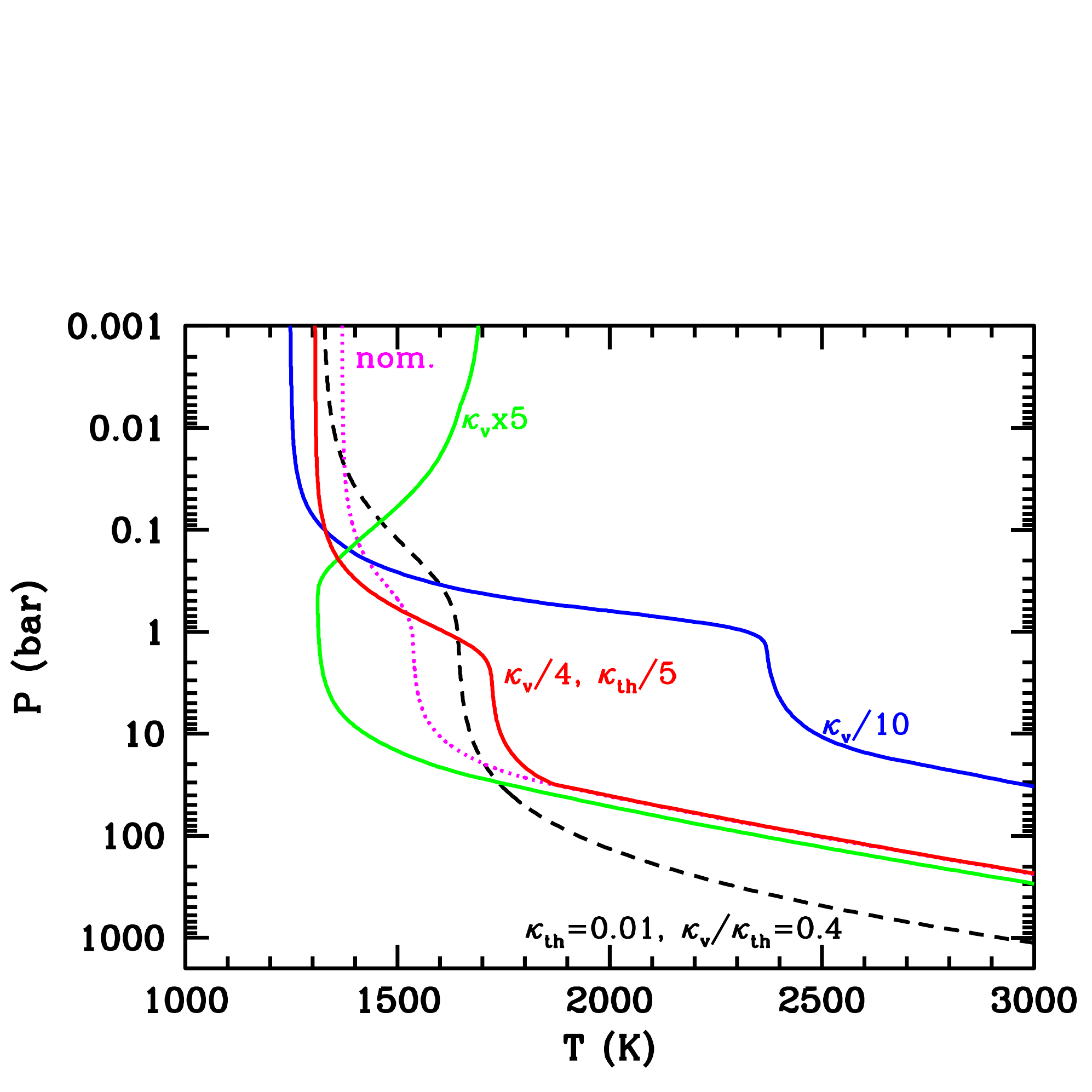}}
\caption{Atmospheric pressure-temperature profiles for a Hot Jupiter found with the  {semi-gray}  approximation of \citet{guillot2010a}. The dotted line is the nominal model, while other lines are calculated with different opacities in the visual and thermal domain.} \label{fig:INTROatmohotjupi} 
\end{center}
\end{figure}

Figure \ref{fig:INTROatmohotjupi} shows a typical example of the atmospheric structure of a Hot Jupiter found with the  {semi-gray} model. The nominal pressure-temperature profile is calculated with the Rosseland mean opacity in the infrared for a solar-composition gas  from \citet{freedmanmarley2008} and the parameter $\gamma$ is set to 0.4. For the black line, the same ratio is used, but the opacity in the thermal domain is fixed to 0.01 cm$^{2}$/g.  A strong reduction of the opacity in the visual leads to a deeper penetration of the irradiation into the planet (blue line), while an enhanced optical opacity leads to a hotter outer atmosphere with a temperature inversion, while the deep atmosphere is cooler (green line). 

Besides the  {semi-gray} atmospheres for strongly irradiated planets that are studied with transit observations, the gray atmosphere should also be replaced with a full coupling of  the internal structure calculations with the non-gray atmospheres of,  {e.g.,} \citet{allardhomeier2011} for non (or weakly) irradiated giant planets, following the procedure explained in \citet{chabrierbaraffe1997}. This will not only provide more accurate cooling curves for giant planets observed with direct imaging, but also their magnitudes in the different observational bands instead of the total luminosity only. This will make  it possible to directly compare the predictions from population syntheses with the discoveries of new direct imaging instruments like \textsc{Sphere}  or \textsc{Gpi}. 
 
\subsection{Infall of planetesimals into the protoplanet's envelope}
This sub-model calculates the interaction of planetesimals and the gaseous envelope of the protoplanet during the formation phase \citep{mordasinialibert2006}. A similar  sub-model was described by \citet{podolakpollack1988} for the classical giant planet models of \citet{pollackhubickyj1996}. This sub-model links the accretion of solid (\ref{sect:INTROsolidaccretionrate}) and the envelope structure (\ref{sect:INTROinternalstruct}). Two quantities are the main output of the sub-model: the protoplanet's capture radius $R_{\rm capture}$ for planetesimals that enters the solid accretion rate (Eq. \ref{eq:safronov}), and the radial energy and mass deposition profiles that enter into the calculation of the envelope structure in particular via the equation for the luminosity. It also yields how much high-Z material is deposited in the envelope to enrich it, and how much of it directly reaches the core. This is important in the context of the (atmospheric) composition of planets \citep{fortneymordasini2013}. The sub-model calculating the interaction \citep[see][for a short overview]{mordasinialibert2006}  includes gravity and gas drag, thermal ablation as for shooting stars, and aerodynamical disruption inspired by the destruction of comet Shoemaker-Levy 9 in Jupiter's atmosphere \citep[e.g.,][]{zahnlelow1994,crawfordboslough1995} or the Tunguska event \citep{chybathomas1993}.

To model the interaction, a set of coupled ordinary differential equations are integrated numerically, giving the position in three dimensions $\mathbf{r}$, velocity $\dot{\mathbf{r}}$, mass $M_{\rm pl}$, and radius $R_{\rm pl}$ of the impacting planetesimal as a function of time $t$. Infalling planetesimals are accelerated by the gravity of the protoplanet and slowed down by gas drag. The equation of motion for the planetesimal  is therefore given by (planetocentric reference frame, 2 body approximation)
\beq
M_{\rm pl}\ddot{\mathbf{r}}=-\frac{Gm(r)M_{\rm pl}}{r^2}\cdot\frac{\mathbf{r}}{r} - \frac{1}{2}C_{D}\rho\ \dot{r}^2\frac{\dot{\mathbf{r}}}{\dot{r}}S 
\eeq
where $C_{D}$ is the drag coefficient that can be written as a function of the Reynolds and the Mach number while $\rho$ is the density of the gas through which the planetesimal is plowing. It is obtained from the calculations of the internal structure of the planet as is the mass $m(r)$ of the protoplanet inside of the position of the planetesimal. S is the cross section of the planetesimal, equal to $\pi R_{\rm pl}^{2}$ for a spherical planetesimal. Note that it is only initially assumed that the planetesimal is spherical, later on it can get distorted due to aerodynamic forces.  

As the planetesimal flies through the envelope,  pressure and temperature are increasing. Eventually, two effects can lead to its destruction: thermal ablation and mechanical mass loss. These effects  determine how deep the planetesimal is able to penetrate, thus determining where in the planet's envelope the energy and the debris of the planetesimals are deposited. The mass loss rate due to thermal ablation can be written in its simplest form as \citep{opik1958}
\beq
\frac{d M_{\rm pl}}{dt}=-\frac{1}{2}C_{H}\rho\ \dot{r}^3 S\frac{1}{Q_{abl}}.
\eeq
$Q_{abl}$ is the amount of energy needed to heat body material from its initial temperature to the point where melting or vaporization occurs plus the specific heat needed for this phase change.  The heat transfer coefficient $C_H$ is an a priori unknown function that can vary over many orders of magnitude \citep[][]{svetsovnemtchinov1995}. It depends on the velocity, envelope conditions, flow regime, shape of the body etc. and gives the fraction of the incoming kinetic energy flux of the gas that is available for ablation. Note that in order to compute this fraction, the hydro- and thermodynamic state of the flow surrounding the  impacting planetesimals needs to be known. The most important  flow regime for massive impactors is a hypersonic highly turbulent continuum flow. This means that a strong shock wave forms. The radiation field generated by the shock is the dominant heating source leading to thermal ablation, as temperatures in excess of 30'000 K are reached \citep[e.g.,][]{zahnle1992}. Therefore, in these conditions, the bow shock temperature must be computed by solving the shock jump conditions for a non-ideal gas \citep[cf.][]{chevaliersarazin1994}. 

Besides thermal ablation, mechanical effects can result in a rapid destruction of the planetesimal by fragmenting it into a large number of small pieces that eventually get thermally ablated. Mechanical destruction is of prime importance for massive (km-sized) impactors as was understood in the hydrodynamical simulations for SL-9 \citep[e.g.,][]{svetsovnemtchinov1995}.  The main effect to consider is the large pressure difference between the front of the planetesimal (stagnation point) and the back where the pressure almost vanishes. This pressure difference (if large enough) leads to a lateral spreading of the body \citep[the so-called ``pancake'' model, introduced by][]{zahnle1992} and eventually its disruption by fragmentation. The rate of lateral spreading is given as
\beq
\frac{d^{2} R_{\rm pl}}{dt^{2}}=C_{\rm S}\frac{\rho}{\rho_{\rm b}} \frac{\dot{r}^2}{R_{\rm pl}}.
\eeq
where $C_{\rm S}$ is a coefficient of order unity \citep{chybathomas1993} while $\rho_{\rm b}$ is the mean density of the fluidized impactor.

The impactor acts approximately as a fluid for ram pressure exceeding the internal tensile strength  {by a large factor}. The disruption into many fragments is then due to Rayleigh-Taylor (RT) instabilities that grow due to the deceleration of the front side of the body (a denser fluid) by the shocked gas (a less dense fluid). Such instabilities are seen to develop in  hydrodynamical simulations \citep[e.g.,][]{lowzahnle1994,korycanskyzahnle2000}. \citet{mordasinialibert2006}  describe this process as a fragmentation cascade due to growing RT fingers. In the non-linear stage, the height until which fingers have grown into the body $h_{\rm inst}$ can be estimated as \citep{sharp1984,youngs1989}
\beq
\frac{d^{2} h_{\rm inst}}{dt^{2}}=2 \alpha \left(\frac{\rho_{\rm b}-\rho_{\rm stagn}}{\rho_{\rm b}+\rho_{\rm stagn}} \right) \ddot{r}
\eeq
where the expression in brackets is the Atwood number,  $\alpha$ a parameter $\approx0.1$ \citep{sharp1984,youngs1989} while $\rho_{\rm stagn}$ is the gas density at the stagnation point. The combined actions of flattening and  growth of RT fingers leads to a fast destruction of the impactor in a terminal explosion, similar as in the Tunguska event \citep[][]{chybathomas1993}.

\begin{figure}
\begin{center}  
  \includegraphics[width=1\columnwidth]{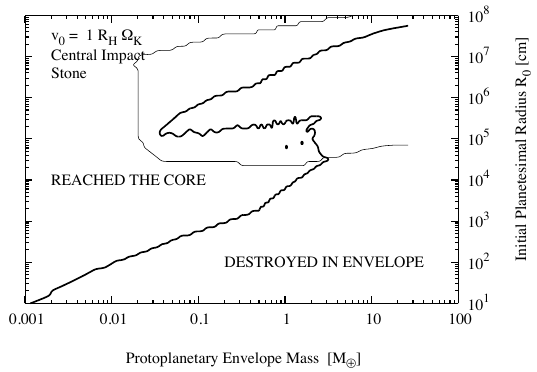}
     \caption{Fate of planetesimals in the envelope of a growing giant planet. The plot shows whether they can penetrate to the solid core or if they get completely destroyed in the envelope as a function of the envelope mass of the protoplanet and the initial radius of the impacting planetesimal  (thick solid line). Planetesimals inside the roughly squared region limited by the thin line in the upper right corner undergo aerodynamic fragmentation (figure from \citealt{mordasinialibert2006}). } \label{fig:INTROinfalling} 
\end{center}
\end{figure}

Figure \ref{fig:INTROinfalling} from \citet{mordasinialibert2006} illustrates the output of the ``infall'' sub-model. It shows the fate of stony planetesimals of various initial radii in protoplanetary envelopes with masses between 0.001 and 100 $\mearth$. The basic result is that the larger the impactor and the lower the envelope mass, the more likely it is that the planetesimal can penetrate to the core, as one expects.  The plot shows that the detailed structure is, however, more complex. There are  intermediate sized planetesimals (radii $\sim$100-1000 meters) that penetrate through surprisingly  massive envelopes. They are too massive for efficient purely thermal ablation, but too small to undergo intense fragmentation. Big bodies ($\gtrsim$100 km) on the other hand are protected by their self-gravity from intense fragmentation.  {It is clear that these results depend on the material properties of the impactors which are typically not well constrained. Icy planetesimal are more prone to destruction in the envelope \citep{podolakpollack1988}, for large impactors mainly due to their lower tensile strength (by about a factor 100, \citealt{chybathomas1993}). If planetesimals have properties similar to the parent body of the Shoemaker-Levy 9 impactors which was a nearly strength-less rubble pile hold together only by self-gravity \citep{asphaugbenz1994}, the ability of planetesimals to reach the solid core would be even more reduced.}

The  {population synthesis framework} shown in Fig. \ref{fig:INTROschema} is currently the only global planet formation and evolution model that explicitly addresses the interaction of planetesimals and the protoplanetary atmosphere. The coupling with the rest of the model is, however, at the moment not yet self-consistent. Despite the fact that  the mass deposition profiles are calculated, it is assumed in the ``envelope'' sub-model that all accreted solids immediately sink to the core \citep[``sinking approximation''][]{pollackhubickyj1996}. The output of the ``infall'' sub-model is thus not used to calculate the (radially changing) composition of the gas, and linked to that, its opacity.  Both are known to be important for the formation process \citep[e.g.,][]{horiikoma2011,movshovitzbodenheimer2010}, and compositional gradients  even have the potential to lead to a semi-convective interior instead of the usually assumed fully convective state in giant planets, with important consequences for the cooling and inferred bulk composition \citep{lecontechabrier2012}.  Including these effects in future work along the lines demonstrated by, e.g., \citet{horiikoma2011} and \citet{iaroslavitzpodolak2007} is therefore important, especially for a more accurate theoretical characterizations of planets in terms of their bulk and atmospheric composition, a quantity that can potentially be measured by spectroscopy \citep[e.g,][]{beandesert2011,konopackybarman2013,bonnefoyboccaletti2013,fortneymordasini2013}. 

\subsection{Internal structure of the solid core}
This  sub-model calculates the radius of the solid core as a function of its mass, bulk composition (iron,  {silicate}, and ice mass fraction), and external pressure due to the surrounding gaseous envelope which is important for giant planets. It is necessary for the correct prediction of the total radius of solid planets without a sizable H/He envelope. But also for giant planets, it is necessary to calculate the core radius correctly to obtain accurate total radii  during the evolutionary phase \citep[e.g.,][]{mordasinialibert2012a}. 

The sub-model for the solid core used in the global model of Fig. \ref{fig:INTROschema} was originally developed in \citet{figueirapont2009a} and assumes a differentiated planet consisting of concentric shells of iron, silicates, and if the planet accreted outside of the iceline, ices. As described in \citet{fortneymarley2007a}, the radius is found by solving the 1D internal structure equations that are in principle the same as for the gaseous envelope. The situation is, however, considerably simplified by assuming that the density of the solid material $\rho$ is (in contrast to  gas) approximately independent of temperature, so that the density is function of pressure $P$ only. The system of equations to solve is then just 
\begin{align}
\frac{d m}{d r}&=4 \pi r^{2} \rho &\frac{d P}{dr}&=-\frac{G m}{r^{2}} \rho
\end{align}
which are the equations of mass conservation and hydrostatic equilibrium ($m$ denotes the core mass inside a radius $r$, $G$ is the gravitational constant).

While \citet{figueirapont2009a} used the more accurate tabulated equations of state (EoS) of  \citet{fortneymarley2007a}, the global model of \citet{mordasinialibert2012} employs the simpler, but more widely applicable EoS of \citet{seagerkuchner2007} that is a modified polytropic EoS  {with the} material parameters  $\rho_{0}$, $c$, and $n$
\beq
\rho(P)=\rho_{0}+c P^{n}.
\eeq
For silicates, the parameters appropriate for perovskite (MgSiO$_{3}$) are used. An advantage of this EoS is that it approaches at sufficiently high pressures (giant planet cores) the correct asymptotic limit which is the EoS of a completely degenerate, non-relativistic electron gas \citep{zapolskysalpeter1969}. The mean density of a core of a massive giant planet can in principle reach very high values exceeding 100 g/cm$^{3}$, with potentially observable consequences \citep{charpinetfontaine2011}. 

\begin{figure}\begin{center}  
{\includegraphics[width=0.99\columnwidth]{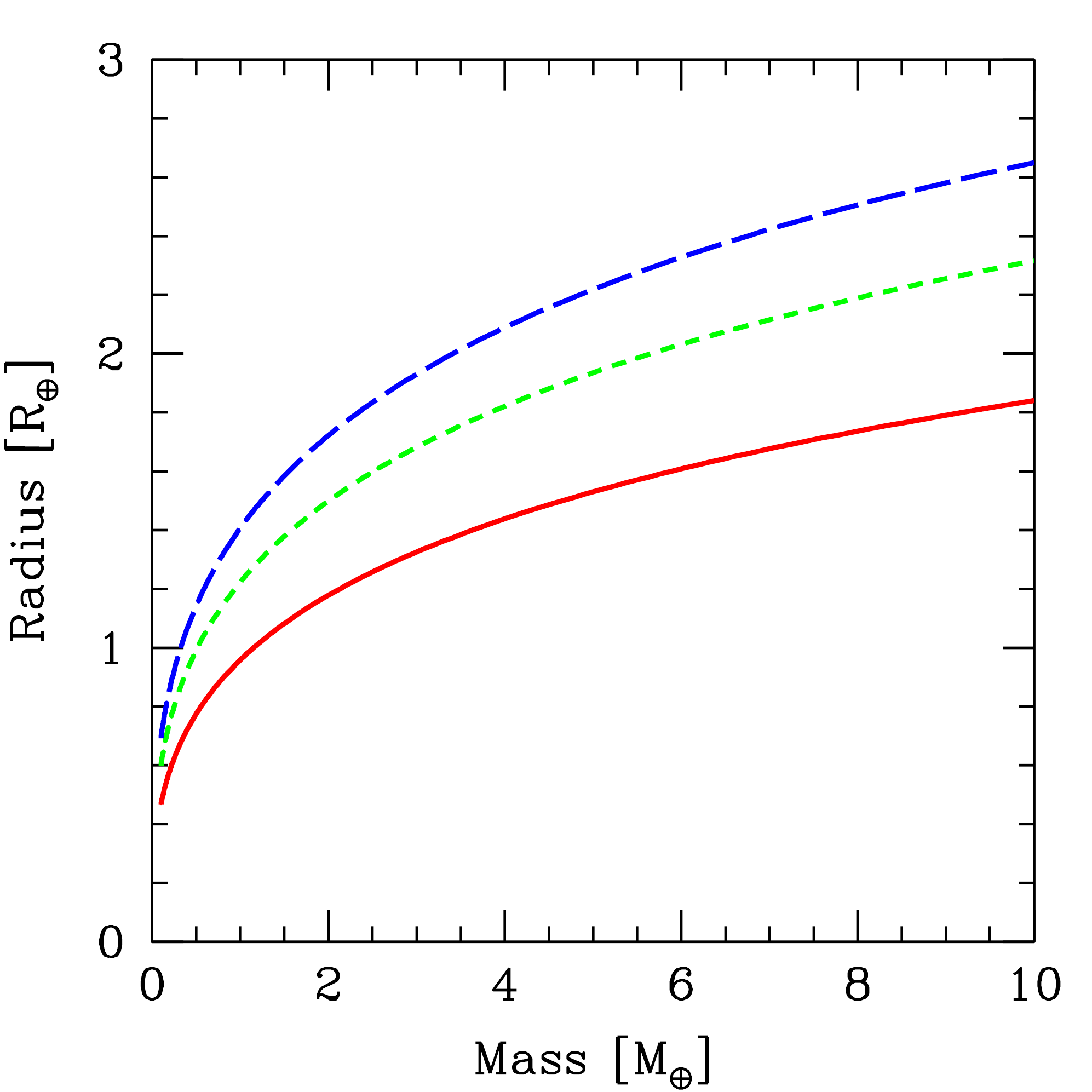}}
\caption{Radius of low-mass solid planets as a function of core mass and composition. The lines show planets having an Earth-like composition (red solid line), pure water ice planets (blue dashed), and a mixed composition with  50\% water ice, 33\% silicates and 17\% iron (green dotted line).  } \label{fig:INTROradidenssolid} 
\end{center}
\end{figure}

Figure \ref{fig:INTROradidenssolid} illustrates the output of this sub-model by showing the radius  as a function of mass for  low-mass solid planets (no external pressure). Three different bulk compositions are shown: Earth-like with a 2:1 silicate-to-iron ratio, pure water ice, and a mixture of both types of materials. The plot shows that the radius follows to good approximation a power-law, as noted by \citet{valenciaoconnell2006}.

A second output of this sub-model is the time-dependent radiogenic luminosity of the solid core that is due to the decay of long-lived radionuclides in the mantle. As described in \citet{mordasinialibert2012}  the radiogenic luminosity can be modeled according to the law of radioactive decay assuming that the mantle has a chondritic composition \citep{urey1955,lowrie2007}. The radiogenic core luminosity is very small compared to the luminosity due to the cooling of the gaseous envelope of a giant planet. For core dominated planets it can, however, become the dominant internal heat source during the evolution over gigayears, and therefore affect the contraction of the planet \citep[e.g.,][]{nettelmannfortney2011}.   {M}uch can be improved  {in the basic description  {of the interior of solid planets} presented here}, like the inclusion of a more accurate EoS, the calculation of the radial temperature structure or the thermal cooling of the  { interior \citep[e.g.,][]{valenciaoconnell2006,lopezfortney2013b}. At early times, the effect of cooling of the core could be particularly important for rocky planets with cores retaining residual heat from the accretion process, especially if the core contains heat from a  violent, short timescale build-up\footnote{ {Note that (giant) planet formation models usually assume that the accretional core luminosity due to planetesimal impacts onto the core is $GM_{\rm c}\dot{M}_{\rm c}/R_{\rm c}$ for a core of mass $M_{\rm c}$, radius $R_{\rm c}$ and planetesimal accretion rate $\dot{M}_{\rm c}$ \citep[e.g.,][]{ricearmitage2003}. This is equivalent to an instantaneous radiation of the entire impact luminosity. In reality, a part of the accretional heating will first be incorporated into the core's interior and then only radiated later on a longer timescale, especially for large impactors \citep{rubienimmo2007}}.}. Also the process of core formation (differentiation of rocky planets into a metallic core and silicate mantle) provides an important heating source, the timescale of which is however not yet entirely clear \citep{rubienimmo2007}. Further improvements could be to include} the dissolution of the solid core in giant planets \citep[e.g.,][]{guillotstevenson2004,wilsonmilitzer2012} and the outgassing of secondary atmospheres given the core's composition acquired during formation \citep[e.g.,][]{elkins-tantonseager2008}. The latter point is  particularly relevant as the James Webb Space Telescope should make it possible to characterize the atmospheres of a few low-mass planets \citep[e.g.,][]{beluselsis2011}.

\subsection{Atmospheric escape}\label{sect:INTROescape}
The \textsc{Kepler} satellite has discovered a very large population of close-in low-mass planets, in agreement with result from high-precision radial velocity searches. Due to their proximity to the host star and their low surface gravity, these low-mass planets are sensitive to envelope mass loss due to atmospheric escape of the primordial H/He envelope \citep[e.g.,][]{lammerselsis2003,baraffeselsis2004,erkaevkulikov2007a,murray-claychiang2009,owenjackson2012a}. This means that  for such planets, atmospheric escape is important on a population level, shaping the statistical properties like the radius distribution \citep{lopezfortney2012,lopezfortney2013,owenwu2013}.  This is due to the fact that the presence of even a very tenuous H/He envelope (low mass fraction)  has a large impact on the total radius \citep{adamsseager2008,rogersbodenheimer2011a,mordasinialibert2012a}.

Therefore, if one wants to connect predictions of a formation model (which yields the H/He mass after formation) with observations by the \textsc{Kepler} satellite, it is necessary to include envelope evaporation during the evolutionary phase. In principle, to find the escape rate, it is necessary to solve the radiation-hydrodynamic  equations describing the flow of the upper atmosphere under the effects of heating by UV and X-ray irradiation on a (potentially multidimensional) grid \citep{murray-claychiang2009,owenjackson2012a}.

A simplified description of the mass loss rate of close-in planets suitable for global models consists of the following elements: First a description for the incoming stellar EUV (and X-ray) flux as a function of time $t$ and orbital distance of the planet $a$ is needed. The EUV flux can be approximated as \citep{ribasguinan2005}
\beq
F_{\rm UV}\approx F_{\rm UV,0}\left(\frac{t}{ \rm 1 \ Gyr }\right)^{-1}\left(\frac{a}{\rm 1 \ AU }\right)^{-2}
\eeq
where $F_{\rm UV,0}$ depends on the host star  type \citep{etangs2007}, while the power-law index for the temporal decrease depends  somewhat on the wavelength interval that is considered.  This equation is valid for stars older than $\sim$100 Myr, while at younger ages, the flux saturates at a maximum value. 

Envelope evaporation can either be driven by X-rays or EUV, depending on the relative position of the ionization front and the X-ray sonic point, which allows to identify which mechanism is dominant \citep{owenjackson2012a}. 

In the EUV regime, two sub-regimes exist \citep{murray-claychiang2009}. At lower EUV fluxes, most of the incoming  energy flux goes into $p dV$ work that lifts gas out of the planet's potential well, while radiative losses and internal energy changes are small. Therefore one can write the evaporation rate with an equation of the form (so-called energy-limited approximation,  \citealt{watsondonahue1981})
\beq
\frac{d M_{\rm UV,e-lim}}{d t}=\frac{\epsilon_{\rm UV} \pi F_{\rm UV} R_{\rm UV}^{3}}{G M K_{\rm tide}}.
\eeq 
In this equation, $M$ is the planet's mass, $\epsilon_{\rm UV}$ is an efficiency factor (that hides the complex physics), $R_{\rm UV}$ the planetary radius where EUV radiation is typically absorbed \citep[estimated as in][]{murray-claychiang2009}, and $K_{\rm tide}$ is a factor to take into account that mass only needs to reach the Hill sphere to escape \citep{erkaevkulikov2007a}.

At high EUV fluxes, most of the UV heating is lost via radiative cooling, so that an equilibrium between photoionization and radiative recombination is established. The evaporation rate in this radiation-recombination limited regime can be approximated as \citep{murray-claychiang2009}
\beq
\frac{d M_{\rm UV,rr-lim}}{d t}=4 \pi \rho_{\rm s} c_{\rm s} r_{\rm s}^{2}
\eeq
where $\rho_{\rm s}$ and  $c_{\rm s}$ are the density and speed of sound at the sonic point at a radius $r_{\rm s}$. These quantities can be estimated as described in \citet{murray-claychiang2009}. 

The mass loss rate in the X-ray driven regime can (roughly) be estimated with an analogous equation as in the energy limited UV regime. These equations are then solved together with the equations for the internal structure and the atmosphere, yielding the evolution of the planet's envelope mass and radius as a function of time. When coupled with population synthesis calculations, this allows to study the global effects of envelope evaporation, for example, on the planetary radius distribution \citep{jinmordasini2014}.

It should, however, be noted that the efficiency factors (both for EUV and X-ray driven evaporation) are in reality not constants, but depend on the planet mass, radius, and magnitude of the heating flux. This means that the mass loss rates found with constant factors should be seen as rough estimates, in particular for the mass loss history of an individual object \citep{owenwu2013}. On the other hand, calculations of the evolution of an entire population of planets under the influence of atmospheric escape indicate that the statistical consequences do not vary drastically when the efficiency factors are varied within reasonable limits \citep{jinmordasini2014}.

\subsection{Orbital migration}\label{sect:INTROorbitalmig}
The discovery of a Jovian planet at an orbital distance of only 0.05 AU from its star by \citet{mayorqueloz1995} was a surprise because theoretical planet formation models had rather predicted \citep[e.g.,][]{boss1995} that giant planets should be found several AU away, as it is the case in the Solar System. The mechanism that was underestimated was orbital migration, i.e., the radial displacement of planets. Several mechanism can cause orbital migration including classical disk migration \citep{goldreichtremaine1980}, migration due to the planetesimal disk \citep[e.g.,][]{levisonthommes2010,ormelida2012}, or Kozai migration with tidal circularization \citep{kozai1962,fabryckytremaine2007}.  {Another possible mechanism producing close-in planets is planet-planet scattering followed by tidal circularization  { \citep[e.g.,][]{ivanovpapaloizou2004,papaloizouterquem2006,beaugenesvorny2012}.} Support for such a dynamical origin of Hot Jupiters arises from the distribution of the orbital distances of these planets relative to the Roche limit   {(\citealt{fordrasio2006,valsecchirasio2014}, but see \citealt{riceveljanoski2012} for a contrasting view)} or the observation of highly inclined planetary orbits relative to the stellar equator via measurements of the Rossiter-McLaughlin effect \citep[see, e.g.,][]{winnjohnson2007,triaudcollier2010}. Such misaligned orbits are not expected from classical disk migration (but see also \citealt{batygin2012}). The relative importance of the various mechanisms that lead to Hot Jupiters is currently debated. A recent comparison of the distribution of the spin-orbit angles indicates \citep{cridabatygin2014} that both disk migration and dynamical mechanisms contribute.}

Here we concentrate on classical disk migration which is the consequence of the gravitational interaction of the gaseous protoplanetary disk and embedded planets.  {This mechanism was the first to be included in most global planet formation models. The dynamical effects in multi-body systems that can potentially also lead to Hot Jupiters were in contrast only addressed  recently in population syntheses \citep{idalin2013,alibertcarron2013}}. The main result of the large body of studies addressing disk migration \citep[see][for a recent review]{kleynelson2012} is that angular  momentum is being transferred between disk and planet via torques that lead in most cases to a loss of angular momentum for the planet (inward migration). The angular momentum $J$ of a planet of mass $M$ in orbit around a star of mass $\mstar$ at a semimajor axis $a$, and the migration rate $d a/dt$ under the action of a total torque $\Gamma_{\rm tot}$ are given as
\begin{align}
    J&=M \sqrt{G \mstar a} &  \frac{d a}{dt}&=2 a \frac{\Gamma_{\rm tot}}{J}
\end{align}

Two types of migration are distinguished depending upon the mass of the planet, respectively its impact on the disk structure. Low-mass planets ($M\lesssim$10-100 $\mearth$) lead to angular momentum fluxes that are much smaller than the background viscous angular momentum transport and therefore do not affect the surface density of the disk in a significant way. They migrate in type I migration where the total torque is found by summing up the contributions from the inner and outer Lindblad torques plus the corotation torque \citep[e.g.,][]{ward1986,tanakatakeuchi2002}. The total torque $\Gamma_{\rm tot}$ can be expressed in a form like \citep{paardekooperbaruteau2010}
\begin{align}
    \Gamma_{\rm tot}&=\frac{1}{\gamma}(C_{0}+C_{1}p_{\rm \Sigma}+C_{2}p_{\rm T}) \Gamma_{0} & \Gamma_{0} & =\left(\frac{q}{h}\right)^{2}\Sigma a^{4}\Omega^{2 }
\end{align}
where $\gamma$ is the adiabatic index of the gas, $q=M/\mstar$, $h$ the disk aspect ratio, $\Sigma$ the gas surface density at the planet's position, and $\Omega$ the Keplerian frequency. The constants $C_{i}$ depend on the local thermodynamical regime in the disk. The quantities $p_{\rm \Sigma}$ and $p_{\rm T}$ are the local radial power-law slopes of the gas surface density and temperature profile in the protoplanetary disk. The  work of \citet{tanakatakeuchi2002} assumed a locally isothermal disk which results in fast inward migration for typical disk conditions. These equations were used in several older population synthesis calculations \citep[e.g.,][]{idalin2008c,thommesmatsumura2008a,mordasinialibert2009,alibertmordasini2011a} where it was found --not surprisingly-- that the migration rates needed to be reduced by large  factors to bring the synthetic results in better agreement with observations. In the meantime it was understood \citep[e.g.,][]{baruteaumasset2008,casolimasset2009,paardekooperbaruteau2010,kleybitsch2009} that in more realistic  non-isothermal disks, there are several sub-types of type I migration, some of which lead to outward migration. The different sub-types can be identified by the comparison of four characteristic timescales, namely, the U-turn, the viscous, the libration, and the cooling timescale \citep{dittkristmordasini2014}. These more realistic type I descriptions were used in several recent population synthesis simulations \citep{hellarynelson2012,dittkristmordasini2014}.

In each type I sub-regime, the migration rate and direction depend besides the planet's mass on the local radial slopes of the surface density and temperature that are given by the disk model (Sect. \ref{sect:INTROradialdisk}). These slopes change due to opacity transitions which has  the consequence in the adiabatic sub-regime that there are special places in the disk with zero net torque and a negative $ d \Gamma_{\rm tot}/da$. This means that inwards of these positions, migration is directed outwards, and outwards of them, it is directed inwards, so that planets in this convergence zone migrate from both sides towards the migration traps \citep{lyrapaardekooper2010,sandorlyra2011b,kretkelin2012b,dittkristmordasini2014}. 

\begin{figure}
\begin{center}  
 \includegraphics[width=0.98\columnwidth]{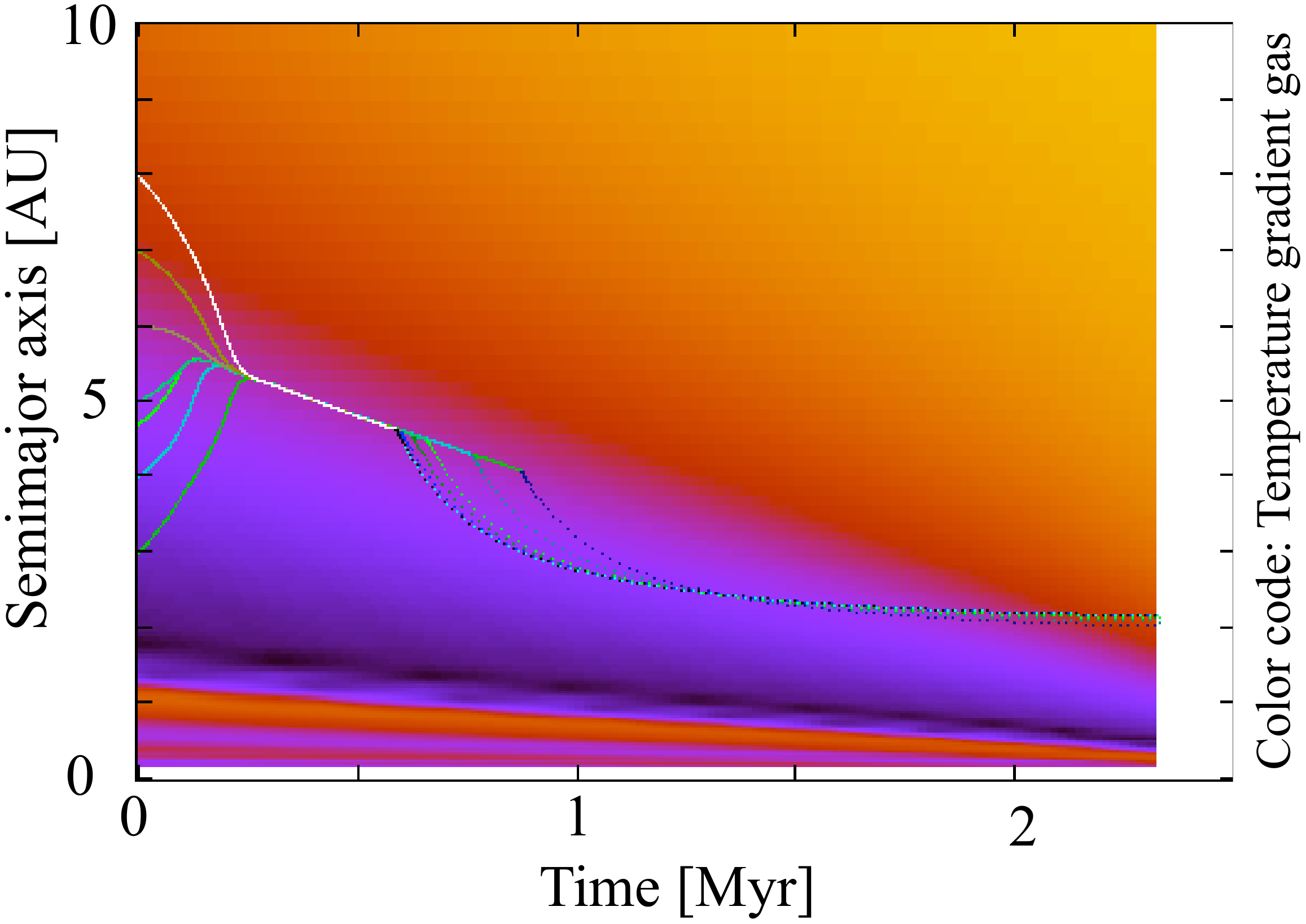}
     \caption{Semimajor axis as a function of time for six migrating and growing  (but mutually non-interacting) protoplanets in the same  viscously evolving disk. The plot shows the rapid migration towards the convergence point and the slow inward migration while the planets are captured into it.   } \label{fig:INTROconvergence} 
\end{center}
\end{figure}

Figure \ref{fig:INTROconvergence} shows the  semimajor axis of six growing planets undergoing non-isothermal migration. All planets migrate in the same protoplanetary disk, but do not interact mutually. The planetary embryos start at different locations ranging from 3 to 8 AU. One sees that at the beginning, the planets starting inside of 6 AU migrate quickly outwards, while those starting further out migrate inwards, so that all planets reach the convergence point where the total torque vanishes. If the disk itself would not evolve, migration would stop at this point.  Due to disk evolution the point of vanishing torque moves itself inwards. This inward migration is much slower than isothermal type I migration as it happens on a viscous timescale \citep{paardekooperbaruteau2010}. During this time, the planets grow by accreting planetesimals and gas.

After a few 10$^{5}$ years, the planets leave the convergence point because the corotation torque saturates, so that they are back at faster inward migration. Shortly afterwards, the planets are sufficiently massive that they transition into type II migration (see below) which is again slower. The planets eventually stop as the local disk mass becomes smaller than the planet mass, so that the inertia of the planet prevents further rapid migration. Simulations that include the gravitational interaction and collisions between the protoplanets find that the convergence zone can be the place of rapid growth as it concentrates a lot of matter in one region. It has therefore the potential to be a preferred place for rapid core formation for (giant) planets \citep[e.g.,][]{hornlyra2012}.

The second main type of disk driven migration is type II migration of sufficiently massive planets. The gravitational interaction of  such massive planets with the protoplanetary disks repels the gas in an annulus around the planet, so that a gap forms \citep{linpapaloizou1986a}. According to the so-called viscous criterion for gap formation, the torque due to the presence of the planet (that repels the gas) must be larger than the background torque in the disk due to turbulent viscosity (that tries to fill up the gap). The additional thermal criterion demands that the disk vertical scale height should be smaller than the planetary Hill sphere radius, so that no sharp, unstable density gradients arise. \citet{cridamorbidelli2006} derived a combined criterion that was used in \citet[][]{mordasinialibert2012} while earlier population syntheses  {partially} only used the thermal criterion.

Type II migration itself comes in two sub-regimes \citep{armitage2007a}. Disk-dominated type II migration occurs if the mass of the planet $M$ is much smaller than the local disk mass ($\sim \Sigma a^{2}$). The planet then acts as a relay that communicates the viscous torques in the disk across the gap by tidal torques. The planet's migration is then locked to the evolution of the disk itself. The migration rate is  {thus} equal to the radial velocity of the gas $v_{\rm r,gas}$ 
\beqa
\frac{d a}{dt}=v_{\rm r,gas}&=&-\frac{3}{\Sigma \sqrt{r}}\frac{\partial}{\partial r} \left(\Sigma \nu\sqrt{r}\right)\\
&=&-\frac{3 \nu}{2 r}-\frac{3}{\Sigma}\frac{\partial}{\partial r} \left(\Sigma \nu \right),
\eeqa
 {but note that hydrodynamic simulations find a more complex behavior depending on the planet mass and gas surface density \citep{edgar2007}.} This means that inside of the radius of maximum viscous couple (or velocity reversal, \citealt{lynden-bellpringle1974}), the planet migrates inwards (which is the normal case) while in the outer parts where the disk is spreading, it moves outwards.

Massive planets in the inner disk (or at late times into the disk evolution) are more massive than the local disk mass, so that planet-dominated migration occurs. The migration rate is then given as
\beq
\frac{da}{dt}=\left(\frac{2 \Sigma a^2}{M}\right)^{k_p}v_{r,gas}
\eeq
where $k_p$ is equal to 1 and 1/2 in the fully and partially suppressed case \citep{alexanderarmitage2009}. The resulting slow-down of planets is important for the final semimajor axis distribution of giant planets \citep{mordasinialibert2009}.

The rough approximation to estimate the planetary type II migration rate based simply on the radial velocity of the gas can be replaced  by the direct summation of the torques according to the original impulse approximation \citep{linpapaloizou1986a,alexanderarmitage2009}. On a longer timescale, it might be desirable for global models to transition to hydrodynamic 2D disk models, because they allow to capture phenomena  that are difficult to model in 1D. An example is the outward migration of two giant planets locked into mean motion resonances \citep{massetsnellgrove2001}.

The global formation model of \citet{hellarynelson2012} uses a similar description of non-isothermal type I  migration and type II migration as presented here. Also \citet{thommesmatsumura2008a} use the isothermal type I migration rates of \citet{tanakatakeuchi2002}, but obtain the type II migration rate based on the more accurate impulse approximation \citep{linpapaloizou1986a}. Their disk model is a 1D viscously evolving model, similar as presented in Sect. \ref{sect:INTROradialdisk}. \citet{idalin2008c} study the global effects of type I migration by population synthesis calculations. They use the migration description of \citet{tanakatakeuchi2002} but reduce the rate  by an arbitrary factor $\leq1$, an approach very similar to the one of \citet{mordasinialibert2009}. This is an example how specialized models can be compared with observations thanks to  population synthesis calculations.  

\subsection{Interaction between several (proto-)planets} 
From the oligarchic growth regime \citep[e.g.,][]{idamakino1993} it is expected that throughout the disk massive bodies (``oligarchs'') form with radial spacings equal to a few mutual Hill spheres. For such a concurrent formation of many protoplanets, multiple effects arise due to the interaction of the planets: regarding accretion, the planets compete for the accretion of gas and solids, excite the random velocities of the planetesimals with consequences for the solid accretion rate of neighboring protoplanets, and increase (for insufficient damping by the disk of gas or planetesimals) the eccentricities of the massive bodies which leads to collisions among the protoplanets and/or their ejection. Also the orbital migration is modified, since migrating planets can get caught into mean-motion resonances, which can, for example, cause outward migration of giant planets \citep{massetsnellgrove2001}. The modification of the surface density of the gas disk due to an already formed giant planets affects the migration of subsequently forming cores \citep[e.g.,][]{massetmorbidelli2006}. The population synthesis calculations of \citeauthor{mordasinialibert2009} (\citeyear{mordasinialibert2009}-\citeyear{mordasinialibert2012}) in contrast assumed that only one planet per disk forms. This one-embryo-per-disk approximation is probably the most severe limitation of the first generation of global models, especially for low-mass planets since the observations show \citep[e.g.,][]{mayormarmier2011} that such planets are mostly found in multiple planet systems. Therefore it is likely that the formation of each planet was influenced by the presence of other bodies. For single giant planets, the impact is probably less severe. 

\begin{figure*}
\begin{center}
       \includegraphics[width=0.8\textwidth]{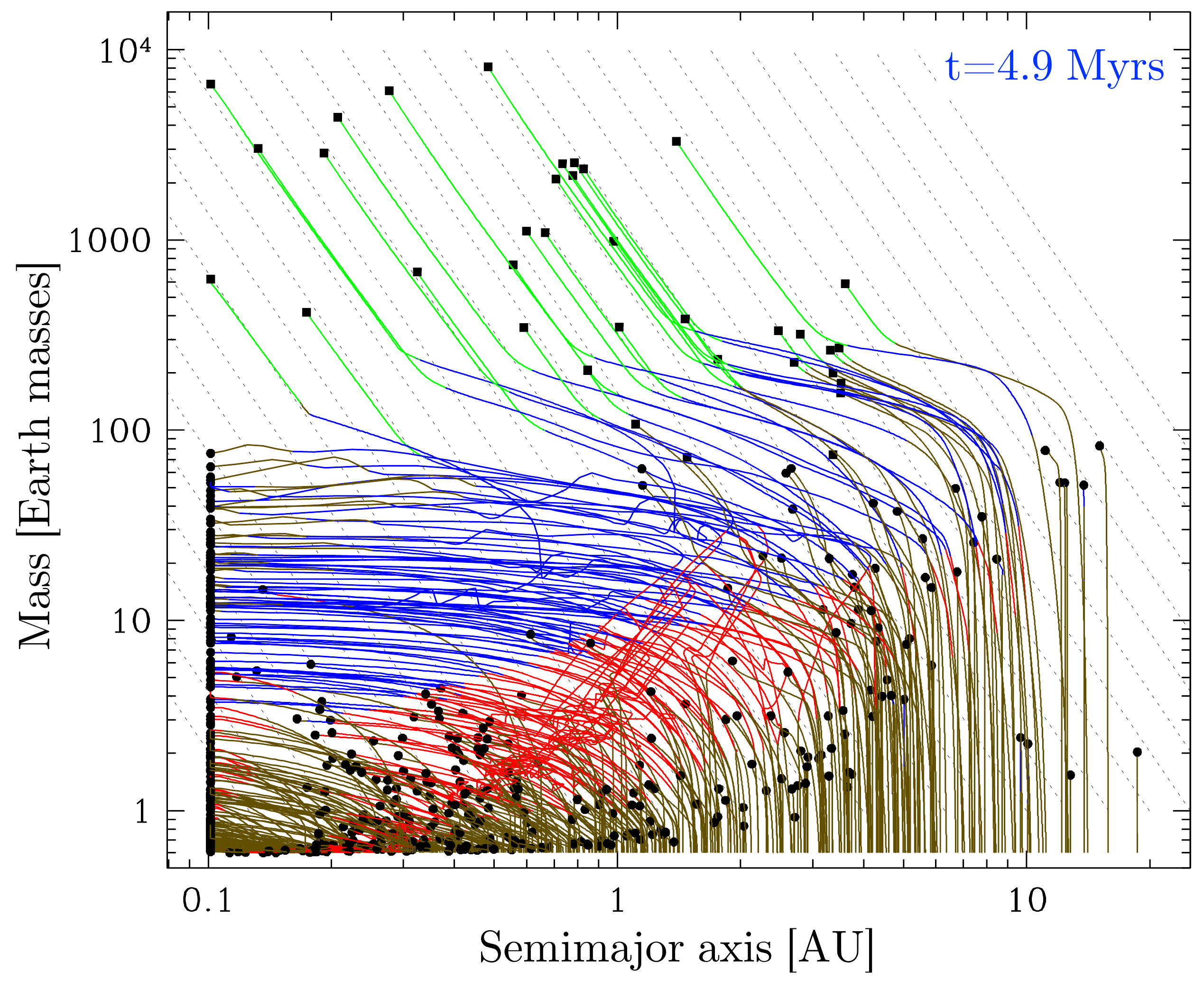}
\caption{Theoretical planetary formation tracks that show how planetary seeds (initial mass 0.6 Earth masses) concurrently grow and migrate. The colors indicate the different types of orbital migration (type I: brown: locally isothermal; red: adiabatic, unsaturated corotation torque; blue: adiabatic, saturated coronation torque; green: type II). The position of the planets at the moment in time that is shown (4.9 Myrs) is indicated by black symbols. Some planets have reached the inner border of the computational disk at 0.1 AU.}\label{fig:INTROtracks}
\end{center}
\end{figure*} 

In the context of global planet formation models, this limitation was recently addressed in \citet{idalin2010,idalin2013}, and \citet{alibertcarron2013}. In the latter work,  an explicit N-body integrator was added ``on top'' of the existing model that calculates the gravitational interaction and the collisions of concurrently forming protoplanets. Typically 10 protoplanets per disk are considered in order to keep the computational time on a level that still allows to conduct planetary population synthesis calculations. The effects of the gravitational interaction with the gaseous disk (eccentricity damping) is modeled as in \citet{foggnelson2007}, and it is assumed that the planetesimal surface density is uniform in overlapping feeding zones. The planetesimals are still described via a surface density and a  mean dynamical state, and not as individual (super-)particles as it is the case in the model of \citet{hellarynelson2012}. In this model, both the protoplanets and the planetesimals (represented by super particles) interact via an explicit N-body integration. The population synthesis models of \citeauthor{idalin2004} (\citeyear{idalin2004a}-\citeyear{idalin2008}) also initially used the one-embryo-per-disk approximation but allowed for several generations of subsequently forming planets. In \citet{idalin2010} and \citet{idalin2013} a new semi-analytical approach was presented to describe the gravitational interaction of several protoplanets in a statistical way based on orbit crossing timescales. The advantage of such an approach is the small computational cost that is several orders of magnitude lower than direct N-body integrations.   Here, one can  clearly see the different origins of various global formation models that are used for statistical studies, like giant planet formation as in \citet{pollackhubickyj1996} for the Alibert, Mordasini \& Benz models,  N-body calculations for the \citet{hellarynelson2012} models, or a dedicated code for a rapid calculations of statistical results as in the case of the Ida \& Lin models.  {An overview of the numerous effects that arise from the dynamical interactions of (proto)planets and external perturbers  {can be found in} \citet{daviesadams2013}.} 

\subsection{Illustrative output: formation tracks}
An illustrative output of the  {population synthesis framework} represented in Fig. \ref{fig:INTROschema} is shown in Figure \ref{fig:INTROtracks}. It shows formation tracks in the mass-distance plane for the one-embryo-per-disk approximation. The  non-isothermal type I migration model is used. Planetary embryos are inserted at a given starting distance into protoplanetary disk of varied properties with an initial mass of 0.6 Earth masses. They then grow by accreting planetesimals and gas, and concurrently migrate due to the interaction with the gas disk.  The distribution of the final positions of the planets (at the moment the protoplanetary disk goes away) can be compared with the observed semimajor axis-mass distribution.

\begin{figure*}
\begin{center}
\begin{minipage}{0.48\textwidth}
	      \centering
       \includegraphics[width=0.9\textwidth]{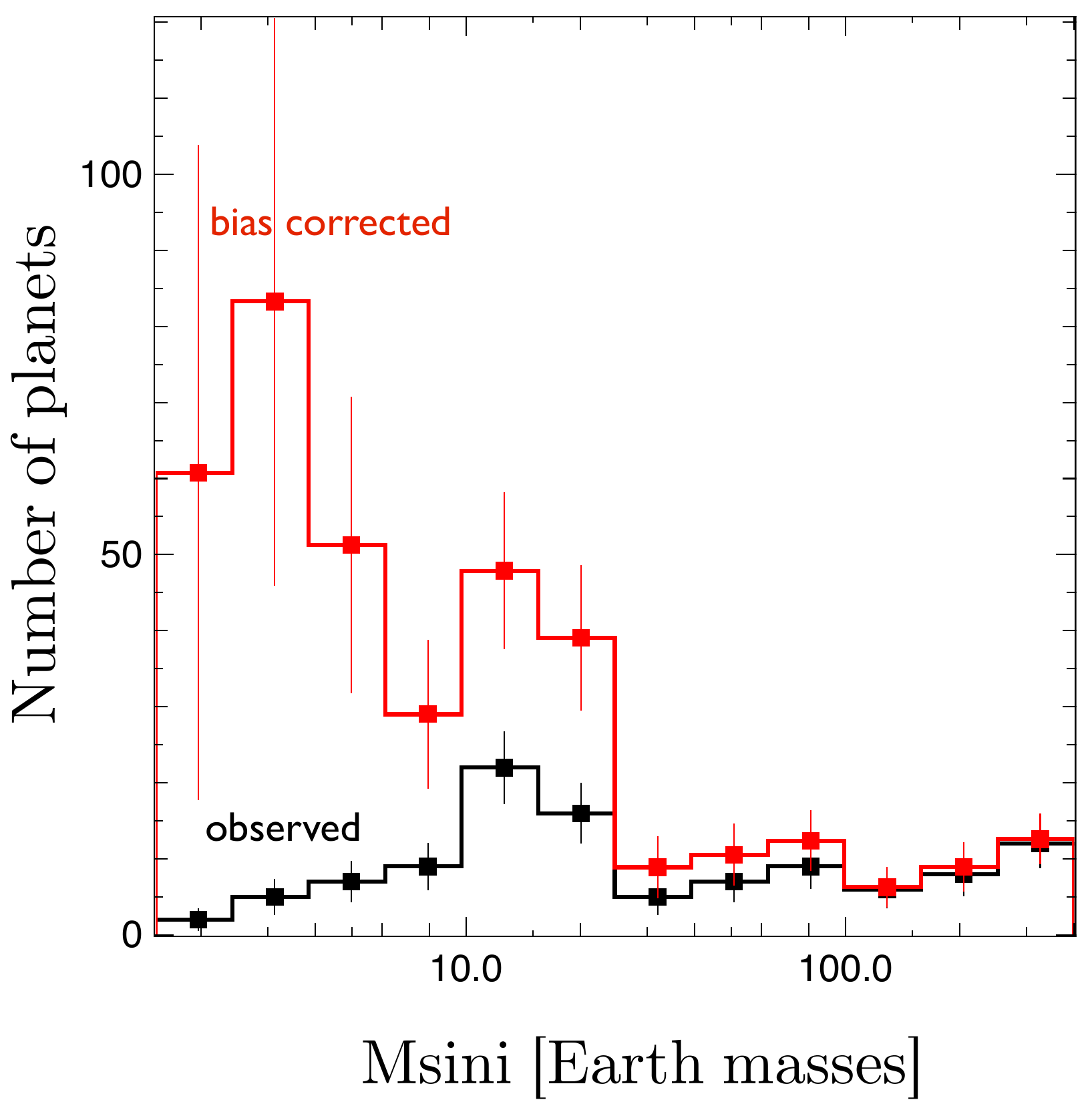}
     \end{minipage}\hfill
     \begin{minipage}{0.5\textwidth}
      \centering
       \includegraphics[width=0.9\textwidth,height=8cm]{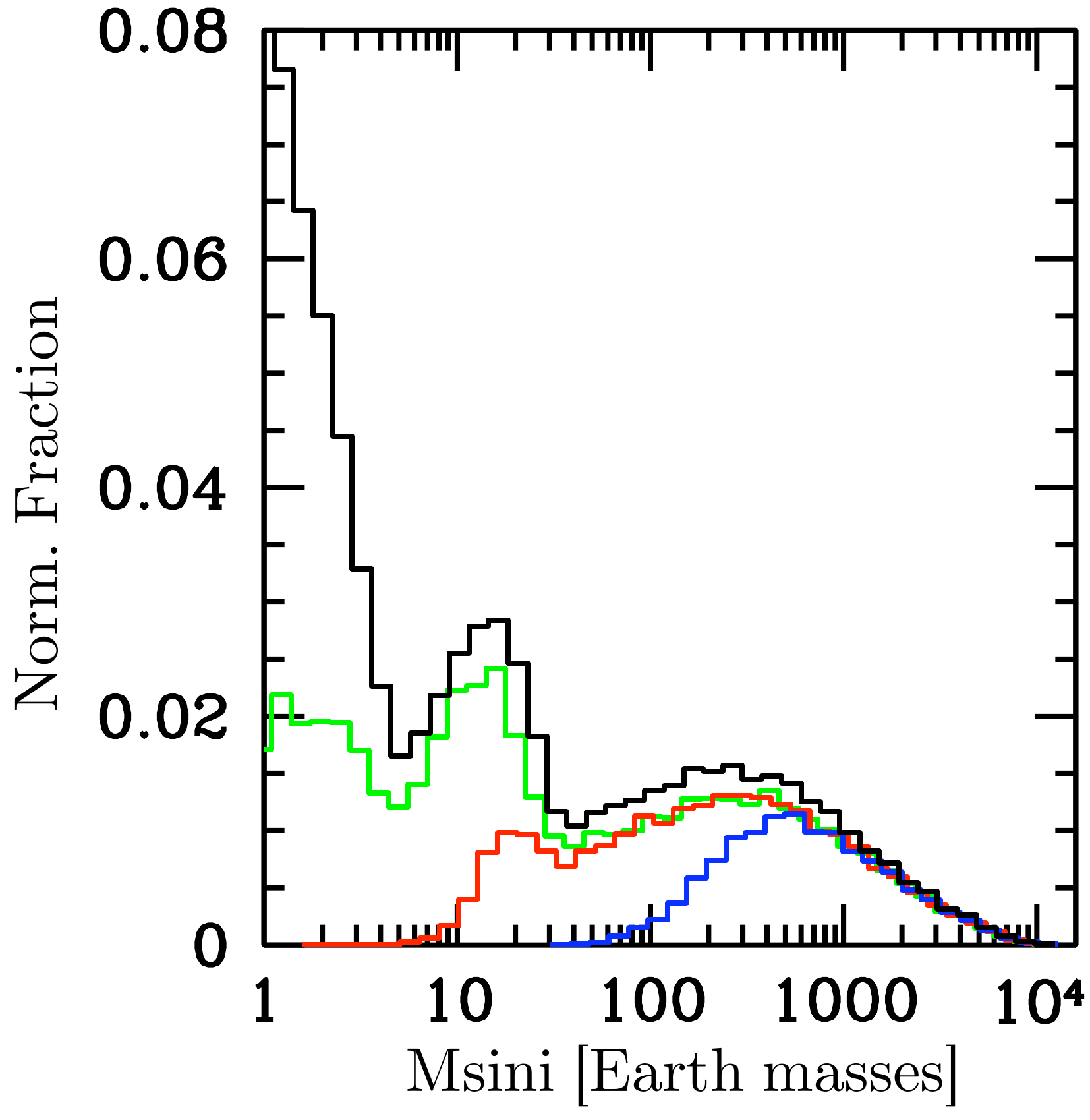}
     \end{minipage}
\caption{Comparison of the observed and the synthetic planetary mass distribution. The left panel shows the distribution of planetary masses as found with high-precision radial velocity observations \citep{mayormarmier2011}. The black line gives the raw count, while the red line corrects for the observational bias against the detection of low-mass planets. The right panel shows the planetary mass function as found in an early population synthesis calculation \citep[figure adapted from][]{mordasinialibert2009}. The black line gives the full underlying population, while the blue, red, and green lines are the detectable synthetic planets at a low (10 m/s), high (1 m/s), and very high  (0.1 m/s) radial velocity precision.}\label{fig:INTROmhisto}
\end{center}
\end{figure*} 

One can see that the outcome of the formation process is of a high diversity, despite the fact that always exactly the same formation model is used. This is a basic outcome similar to the observational result. In the figure, one can, for example, find tracks that lead to the formation of hot Jupiters. Most embryos, however, remain at a low mass, since they cannot accrete a sufficient amount of planetesimals to start rapid gas accretion and become a giant planet. At an orbital distance of 0.2 to 1 AU, an over-density of low-mass planets ($M\lesssim5 \mearth$) can be seen. These are planets that are captured in the inner convergence zone (cf. Fig. \ref{fig:INTROconvergence}). One also notes that  almost all giant planets are inside of 1 AU, which is not in agreement with observations. This points to too rapid inward orbital migration in the model, meaning that the theoretical description of this process must be further improved. It is a typical result that the synthetic mass distribution (discussed in the next section) is in better agreement with the observational data than the synthetic semimajor axis distribution \citep[e.g.][]{mordasinialibert2009}.

\section{Comparisons with observation}
In this section we discuss important selected comparisons between theoretical and observed statistical properties. Thanks to the coupling of planet formation and evolution  {in the global model shown in Figure \ref{fig:INTROschema}}, it is possible to compare with all major observational techniques. 

\subsection{Radial velocity: the planetary initial mass function}\label{sect:INTROrvcomp}
Among the many outputs that can be compared with observations, one of the most fundamental results of population synthesis is a prediction for the distribution of planetary masses. It is obvious that the planetary mass function has many important implications, including  the question about the frequency of habitable extrasolar planets. In the left panel of  Fig. \ref{fig:INTROmhisto}, the planetary mass function is shown as derived  from the high-precision radial velocity  \textsc{Harps} survey of FGK dwarfs  \citep{mayormarmier2011}. It makes clear that below a mass of approximately 30 Earth masses, there is a strong increase in the frequency. In particular, low-mass planets of  with masses less than $\sim$10$\mearth$  are very frequent. The right panels shows the predicted mass function from an early population synthesis calculations of planets around a 1 $M_{\odot}$ star (\citealt{mordasinialibert2009} {, see also} \citealt{idalin2004a}). The synthetic model predicted a large population of low-mass planets. A feature that is  very interesting from a theoretical point of view is that also in the theoretical curve, there is a strong change in the frequency at a similar mass as in the observations. This is explained by the fact that in this mass domain (beyond the critical core mass), planets start to accrete nebular gas in a rapid, runaway process \citep{mizunonakazawa1978}. They then quickly grow to masses  $M\gtrsim100\mearth$. It is unlikely that the protoplanetary disk disappears exactly during the short time during which the planet transforms from a Neptunian into a Jovian planet.  This makes that planets with intermediate masses $\gtrsim 30 \mearth$ are less frequent (``planetary desert'', first found in \citealt{idalin2004a}). The ``dryness'' of the desert depends directly on the rate at which planets can accrete gas during the runaway phase (see  \citealt{mordasinimayor2011}), while the mass where the frequency drops represents the mass where runaway gas accretion starts, i.e., the critical core mass which is of order 15$\mearth$, so that the total mass is about 30$\mearth$ (core and envelope mass are approximately equal when the rapid accretion of gas sets in, \citealt{pollackhubickyj1996}). We thus see how the comparison of synthetic and actual mass function constrains the core accretion model. We further note that while qualitatively, model and observation agree in the basic result that low-mass planets are very frequent, quantitatively the number of detectable planet at 1 and 0.1 m/s is  clearly underestimated in the model. This could partially be related to the one-embryo-per-disk approximation that is used in this  {early} simulation (see \citealt{benzida2013} for an updated synthetic mass function).

\subsection{Astrometry and microlensing: exploring different sub-populations}
The observational constraints from astrometric observations are in principle similar to those from the radial velocity technique, with the difference that the actual mass is measured, and that the detection sensitivity increases with  semimajor axis. To date, detecting extrasolar planets with this technique has proven difficult to achieve \citep[e.g.,][]{sahlmannsegransan2011}, but is the ultimate goal of a number of ground based (e.g., \textsc{Prima}, \citealt{launhardtqueloz2008}) and space based missions (like the proposed \textsc{Neat} satellite, \citealt{malbetleger2011}).  In any case, the \textsc{Gaia} satellite is predicted to discover a very large number of extrasolar giant planets  (several thousands at intermediate orbital distances of $\sim$1 to 4 AU, \citealt{caseratnolattanzi2008}). Since these discoveries will result from an unbiased, magnitude-limited survey with a well defined detection bias (similar to the \textsc{Kepler} satellite), they will be extremely useful for statistical studies of giant planet formation.

Also  results from the microlensing technique are important for statistical studies, since they probe the sub-population of low-mass planets at a few AU of orbital distance that is not accessible to other techniques. Microlensing is in some sense an extreme statistics-only method, because in its simplest form it only yields the distance of the planet in units of the star's Einstein radius and the ratio of the planet's mass to the (unknown) host star mass, meaning that very few physical information about an individual planet is revealed (this can change if  a number of effects like the microlensing parallax,  the orbital motion of the planet, or finite source effects can be measured, see, e.g., \citealt{gaudi2012}). 

For statistical studies, this is not necessarily a problem, provided that the number of detections is sufficiently high and that the observational detection bias is well known. A number of studies \citep[e.g.,][]{goulddong2010,cassankubas2012} have already used the microlensing discoveries to derive the frequency and a power-law exponent for the planetary mass function.  From a theoretical point of view it seems rather unlikely that two parameters (normalization and power-law slope) are sufficient to describe the planetary mass function over  three orders of magnitude in mass (from a few $10^{0}$ to a few $10^{3}$ $\mearth$) as it is currently made in the observational studies due to the low number of microlensing planets. From the core accretion paradigm one expects that at least four parameters are needed (this will still be a rough approximation only), because there are two different fundamental  types of planets (solid planets and gas giants) for which different physical mechanisms determine the mass. The two types should therefore come with separate slopes and normalizations as one can already deduce from Fig. \ref{fig:INTROmhisto}. As the number of microlensing planets is expected to increase thanks to ground and space based observations with satellites like \textsc{Euclid} \citep{pennykerins2013} or \textsc{Wfirst} \citep{goullioudcontent2012}, it will become possible to test this prediction observationally. 

\begin{figure}
\begin{center}
       \includegraphics[width=0.999\columnwidth]{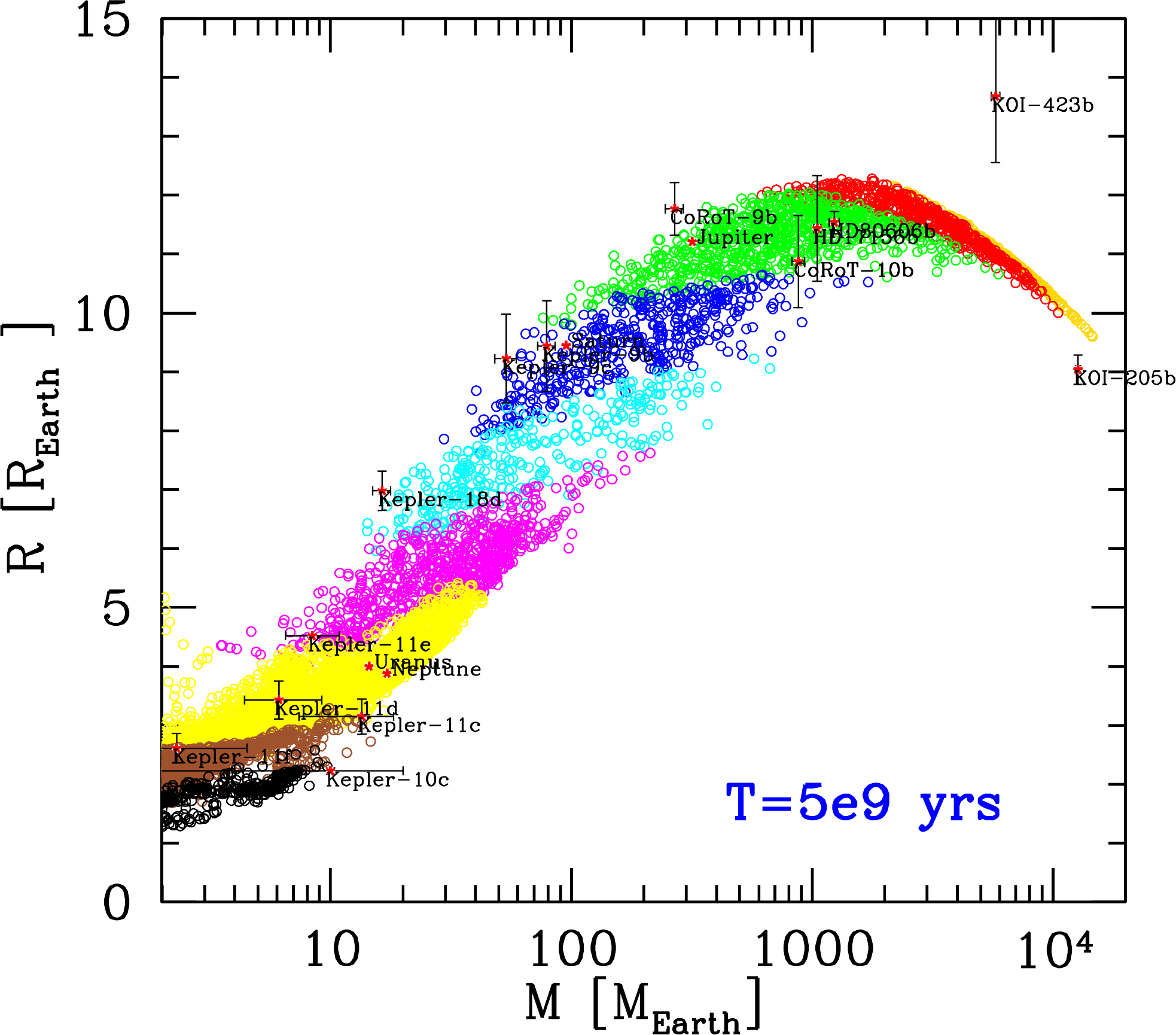}
\caption{Mass-radius diagram of synthetic planets with a primordial H/He envelope at an age of 5 Gyrs and a semimajor axis between 0.1 and 5 AU together with  {actual} planets in- and outside of the Solar System with a  {well} known mass and radius, and a semimajor axis of at least 0.1 AU.  The colors indicate the mass fraction of heavy elements $Z$ in the synthetic planets.  The black symbols, for example, correspond to solid-dominated low-mass planets which contain at most 1\% of H/He, while the most massive planets (dark yellow) consist of at least 99\% H/He. The other colors are:  Red: 1$<Z\leq$5\%. Green: 5 $<Z\leq$20\%. Blue: 20$<Z\leq$40\%. Cyan: 40$<Z\leq$60\%. Magenta: 60$<Z\leq$80\%. Yellow: 80$< Z\leq$95\%. Brown: 95$<Z\leq$99\% \citep[figure modified from][]{mordasinialibert2012}.}\label{fig:INTROmr}
\end{center}
\end{figure} 

\subsection{Transits}\label{sect:transits}
In the past few years, there was a very rapid increase of  observational data coming from photometric observations, in particular  {thanks} to the \textsc{Kepler} satellite. This new observational data provides important new impulses to planet formation and evolution theory, since it adds constraints that go beyond the position of a planet in the mass-distance diagram. Extended comparisons of theoretical and observational results derived from transit observations can be found  {for example in \citet{howardmarcy2012}, \citet{mordasinialibert2012}, \citet{lopezfortney2013b} or \citet{marcyisaacson2014} to name just a few}. Here we concentrate on two important results. 

\subsubsection{Synthetic mass-radius diagram}
The first is the mass-radius diagram. The observed mass-radius relationship was  {shown} in Sect. \ref{sect:obsmassradiusdia}. Here we discuss its synthetic counterpart. Figure \ref{fig:INTROmr}  shows a comparison of the  mass-radius relationship of actual and synthetic planets as found in a  {population synthesis} that combines planet formation and evolution  \citep[modified from][]{mordasinialibert2012}.  The global shape of the planetary mass-radius relationship can be understood from the core accretion paradigm and the basic properties of matter as expressed in the equations of state: Low-mass planets can only accrete tenuous H/He envelopes since their Kelvin-Helmholtz timescale of envelope contraction during the formation phase is long compared to the typical lifetime of a protoplanetary disk \citep[e.g.,][]{ikomanakazawa2000}. Therefore, the top left corner in the $M$-$R$ plane remains empty, as no low-mass, gas-dominated planets come into existence. 

Also the bottom right  {and middle part} remains empty.  {For typical protoplanetary disks, this is simply a consequence of the fact that such disks do not contain enough mass in metals to form a solid Jovian-mass planet.  For the MMSN, for example, one can roughly estimate a total amount of heavy elements of  $M_{\rm disk,MMSN} \times Z_{\rm \odot}\approx 0.013 M_{\odot}\times 0.015$ \citep{hayashi1981,lodders2003} which corresponds to about  64 $\mearth$. However, for a very massive metal rich disk, we can in contrast have a disk gas mass that is still self-gravitationally stable of about 10\%  of the stellar mass and a metallicity [Fe/H]=0.5. Then one estimates for a 1 $M_{\odot}$ star a total mass of heavy elements of order $1 M_{\odot} \times 0.1 \times Z_{\odot} \times 10^{0.5}$ which gives about 1600 $\mearth$ (for a similar discussion, see \citealt{baraffechabrier2008a}). Obviously, it is not clear how much of this mass can be incorporated into one planet and at what point of the formation process this should happen. 

In this context it useful to note that a number of  transiting giant planets seem to have extreme amounts of metals in their interior as estimated from their mass-radius relation: for Hat-P-20b \citet{lecontechabrier2011} estimate  of order 340 $\mearth$ of metals (water) in the interior; about 300-1000 $\mearth$ of solids seem to reside in Corot-20b (\citealt{deleuilbonomo2012}, using the internal structure model of \citealt{guillotmorel1995}). This indicates that large amounts of solids are indeed available in some disks and that some planets even manage to accrete a significant fraction of the totally available metal inventory. From these considerations it appears that just from the availability of metals in protoplanetary disks, it is not excluded that in principle a, say, 100 $\mearth$ planet made entirely of ices could form. Yet, no such planet has been observed to date: all known planets in this mass range ($M\gtrsim100 \mearth$) have a mass-radius relation that shows that they contain significant amounts of H/He (see Fig. \ref{fig:INTROmrobs}).}

This is  {expected from the core accretion model} since massive cores necessarily cause  runaway gas accretion \citep[e.g.,][]{papaloizouterquem1999}, so that the final composition of the forming planet contains  {significant amounts} of envelope gas,  {at least if the cores form during the presence of the gaseous nebula. Thus,} no massive purely solid planets come into existence that would populate the bottom right  {and middle part of the mass-radius  diagram}. One further notes that the synthetic and most actual planets (both in- and outside of the Solar System) populate similar loci in the mass-radius plane.  

From the position of a planet in the mass-radius relationship it is possible to deduce (within limits due to the degeneracies, cf. \citealt{rogersseager2010}) the bulk composition of a planet. The plot shows that depending on the mass range, there can be many different associated radii for one mass, reflecting a large diversity in  interior compositions (in this case, fraction of heavy elements versus H/He). These different compositions are in turn due to the different formation histories. It is, for example, found that for the low planetesimal random velocities assumed here, planets at large distances typically contain a higher fraction of solid elements, since the mass of planetesimals available to accrete (the isolation mass) increases for typical disk models (radial slope of the planetesimal surface density) with distance. This correlation is preserved even under the action of disk migration. It might not be preserved if scattering is responsible for the formation of close-in planets. This could open a new possibility to distinguish different modes of formation for Hot Jupiters. 

\subsubsection{Impact of grain opacity on the planetary radius distribution}\label{sect:grainopacity}
\begin{figure*}
\begin{center}
       \includegraphics[width=1\textwidth]{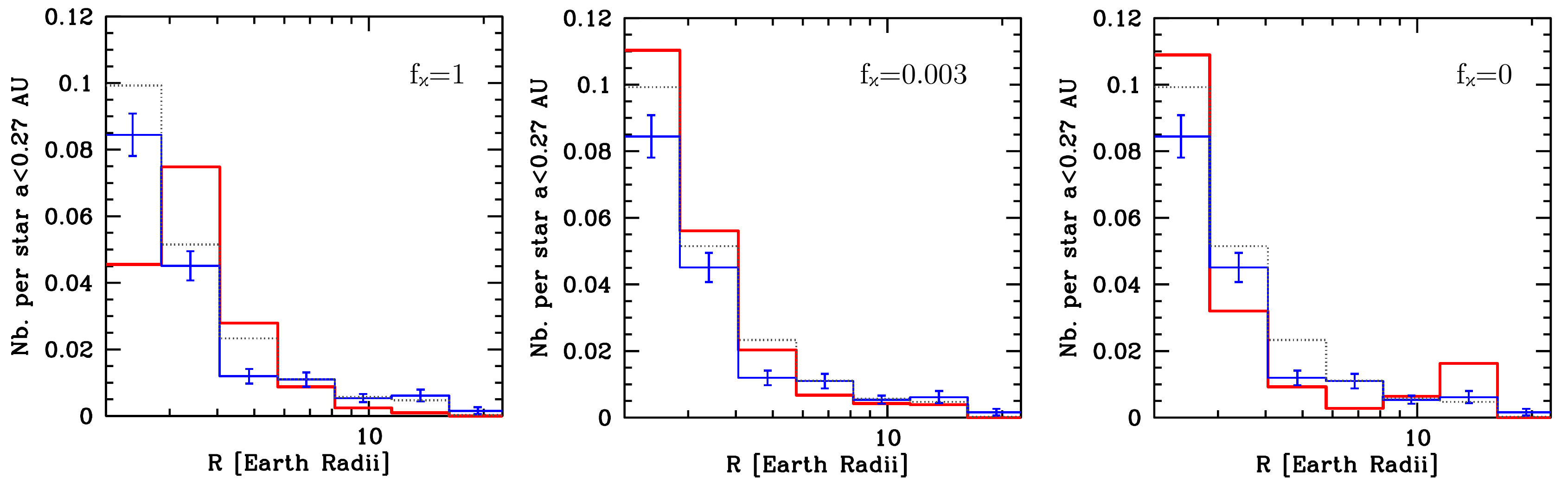}
\caption{Impact of the opacity due to grains in the protoplanetary gas envelope during the formation phase on the planetary radius distribution after 5 Gyrs of evolution. In all panels, the  blue line with error bars shows the bias-corrected distribution of radii found by the \textsc{Kepler} satellite (data set from \citealt{boruckikoch2011b})  according to the analysis of \citet{howardmarcy2012}. The black dotted lines are the preliminary  analysis by the same authors of the updated \textsc{Kepler} data set of \citet{batalharowe2012}. The red line shows the synthetic distribution for an opacity due to grains in the protoplanetary envelope equal to the ISM grain opacity, reduced relative to the ISM by a factor 0.003, and for vanishing grain opacity (left to right). }\label{fig:INTROrhisto}
\end{center}
\end{figure*} 	 
The second example of a synthetic result that can be compared with transit observations is the planetary radius distribution. It can serve as an example of how planet population synthesis can be used to study the global  {consequences} of  a specific physical mechanism  {\citep[see][]{mordasiniklahr2014,hasegawapudritz2014}.}

Figure \ref{fig:INTROrhisto} compares the observed distribution of radii of planets inside of 0.27 AU as found by the \textsc{Kepler} satellite \citep{howardmarcy2012} with the  radius distribution in three different population syntheses. The three calculations are identical except for the value of $f_{\kappa}$. This parameter describes the reduction factor of the opacity due to grains suspended in the protoplanetary atmosphere during the formation phase relative to the interstellar value. A value of $f_{\kappa}=1$ means that the full interstellar opacity is used (\citealt{belllin1994}), while $f_{\kappa}=0$ means that we are dealing with a grain-free gas where only molecular opacities contribute \citep{freedmanmarley2008}. These opacities are used when calculating the internal structure of the planets (Sect. \ref{sect:INTROinternalstruct}) where the  magnitude of the opacity is important for the rate at which planets can accrete primordial H/He envelopes \citep[e.g.,][]{ikomanakazawa2000}. At low opacities, the liberated gravitational potential energy of the  accreted gas can be more promptly radiated away, allowing the envelope to contract faster, so that new gas can be accreted. Specialized microphysical models of grain evolution predict that the opacity should be strongly reduced in protoplanetary atmospheres relative to the ISM because grains grow rapidly in the denser atmosphere and then settle into the deeper parts of the envelope where they are vaporized \citep[][]{podolak2003,movshovitzpodolak2008,movshovitzbodenheimer2010}.

 {Considering} the radius bin in Fig. \ref{fig:INTROrhisto} that contains the giant planets at about 1 Jovian radius ($\approx$11 Earth radii), we see that with full grain opacities, there are too few synthetic planets relative to the observations. With vanishing grain opacities, the efficiency of giant planet formation is on the contrary too high in the model relative to the data.  A relatively good agreement with the observations is found with a $f_{\kappa}=0.003$ (middle panel), which is the value derived from fitting gas accretion timescales found with  detailed grain growth models (\citealt{movshovitzbodenheimer2010}). Even if this comparison is preliminary (it is unclear whether the fitting value found for a specific simulation can generalized to the entire population, see \citealt{mordasiniklahr2014}), it shows that the grain opacity has statistically  {visible} population-wide consequences. This open the possibility to observationally test microphysical grain models.

\subsection{Direct imaging: luminosity at young ages}\label{sect:INTROdirectimaging}
The direct imaging technique measures the intrinsic luminosity of young Jupiters. This is interesting for planet formation and evolution theory because it is an observable quantity that is determined  {by the entropy of  the gas in the interior of the planet. The entropy is in turn determined}  by the structure of the accretion shock of gas during the gas runaway accretion phase.  {Even} more fundamentally,  {the entropy state could be related to} the basic giant planet formation mechanism (core accretion versus gravitational instability, see, e.g., \citealt{spiegelburrows2012,galvagnihayfield2013}).

\begin{figure}
\begin{center}  
{\includegraphics[width=0.99\columnwidth]{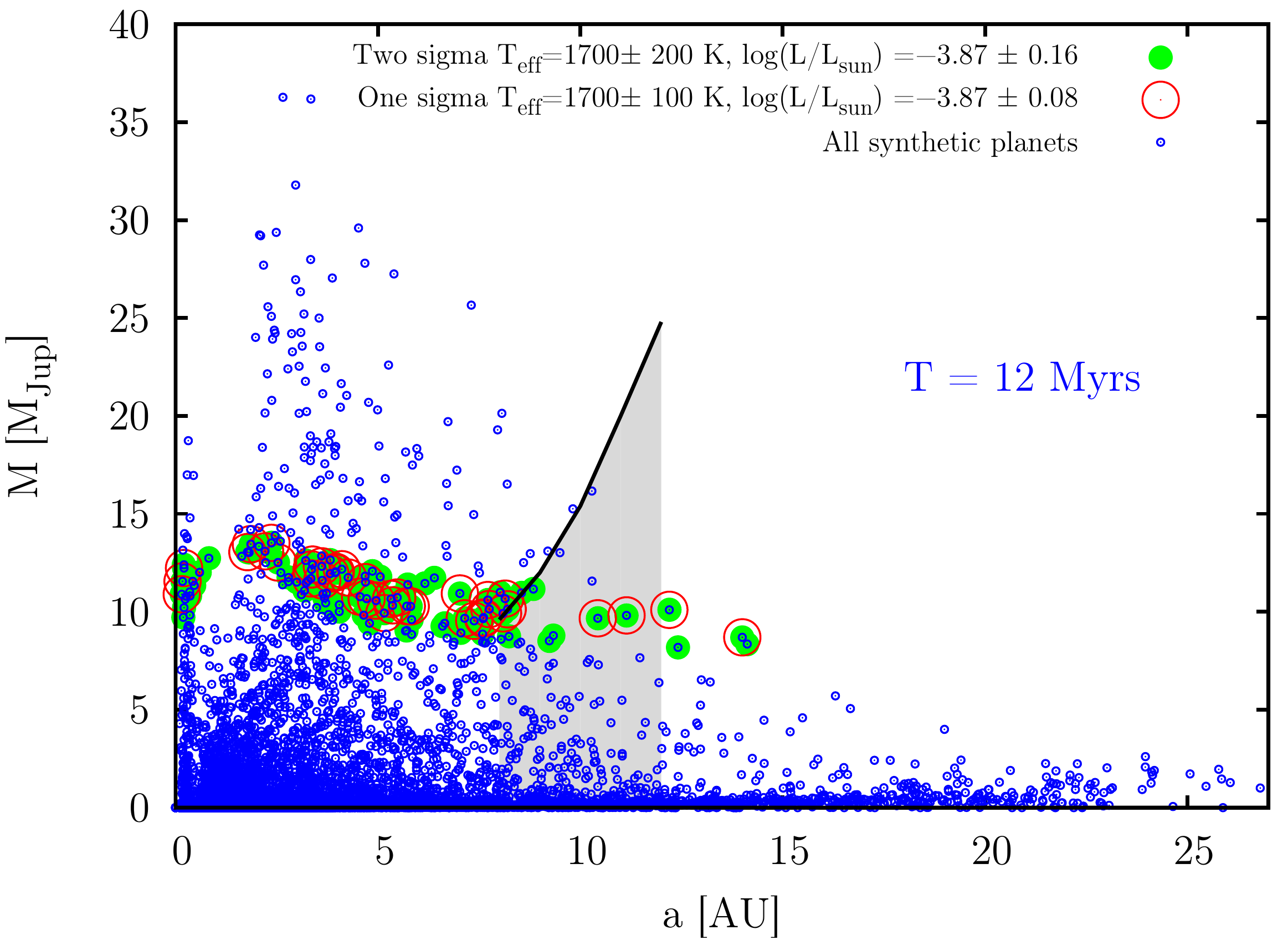}}
\caption{Comparison of theoretical results and  combined observational constraints on the nature and formation of $\beta$ Pictoris b from radial velocity observations and direct imaging \citep[adapted from][]{bonnefoyboccaletti2013}. The plot shows the mass-distance diagram of synthetic planets predicted for the specific properties of the host star. The blue points represent all synthetic planet while the large green circles and red open symbols indicate planets that agree at an age of 12 Myrs within two respectively one $\sigma$ with the constraints derived from direct imaging ($T_{\rm eff}$ and $L$). The  gray shaded region corresponds to the mass domain that agrees with the limits derived from radial velocity observations.  } \label{fig:INTRObetapicb} 
\end{center}
\end{figure}

The number of planets detected by this method has increased significantly in the last few years \citep[see][for an overview]{marleaucumming2013}. In absolute numbers it is still  low compared to the radial velocity and transit method, but this should change in the coming years. It is nevertheless already now possible to use statistical methods to study the discoveries made by this method as shown in Figure \ref{fig:INTRObetapicb}. 

The figure show the mass-distance diagram for a synthetic population that is customized for the star $\beta$ Pictoris which is  a young ($\sim$12 Myr) A5V star visible to the naked eye. Customized means that the quantities that are known for this specific star (its mass and metallicity) are fixed at the observed values, in contrast to normal population syntheses where these quantities are Monte Carlo variables. Additionally, the properties of the synthetic planets are studied at the actual age of the star. 

$\beta$ Pictoris is orbited by a directly imaged companion at a semimajor axis of about 8-10 AU \citep{lagrangegratadour2009,lagrangebonnefoy2010}. The recent analysis of the near-infrared spectral energy distribution by \citet{bonnefoyboccaletti2013} shows that the companion has a luminosity of  $\log(L/L_{\odot})$=-3.87$\pm$0.08, an effective temperature $T_{\rm eff}$=1700$\pm$100 K and a surface gravity $\log(g)$=4.0$\pm$0.5 (cgs units). According to ``hot start'' models \citep[e.g.,][]{burrowsmarley1997,baraffechabrier2003} that assume an arbitrary (high) value of the entropy after formation without considering the actual formation phase, these values correspond to a mass of the companion of 9$^{+3}_{-2}$ Jovian masses. Radial velocity observations \citep{lagrangedebondt2012} show that the mass of the companion must be less than 10 and 25 Jovian masses for semimajor axes of 8 and 12 AU, respectively. The allowed mass domain is indicated in Figure \ref{fig:INTRObetapicb}.

The mass-distance diagram as a function of stellar mass was studied  {in the past} with models that predict only these quantities \citep{idalin2005a,alibertmordasini2011a}. The new result shown in Fig. \ref{fig:INTRObetapicb} is that  {the theoretical} model  predicts besides mass and semimajor axis also the luminosity and  effective temperature.  {Thus} it becomes possible to combine the constraints from both direct imaging and radial velocity.  The plot shows that there are indeed a number of planets that agree in all four fundamental properties ($a$, $M$, $L$, and $T_{\rm eff}$) with the observed values, indicating that $\beta$ Pictoris $b$ could have formed by core accretion and that it has a mass in the planetary mass domain of about 10 Jovian masses. Note that this result directly depends on the assumption in the formation model that the planetesimal random velocities are low as in \citet{pollackhubickyj1996} and that gap formation does not lead to a reduction of the gas accretion rate \citep{kleydirksen2006}.

The simulations are  {conducted} assuming that during gas runaway accretion, the gas accretion shock radiates all potential energy liberated by the infalling gas (``cold accretion''). Still there are planets that have a luminosity that agrees with the observed value of about $\log(L/L_{\odot})$=-3.9. This might seems surprising at first, because under the same assumption, \citet{marleyfortney2007} had in contrast found that the post-formation luminosities in the relevant mass domain are always less than $\log(L/L_{\odot})$=-5. While investigating the reason for the discrepancy, it was found that the difference between the simulations stems from different core masses $\mcore$: in the \citet{marleyfortney2007} simulations, the core masses are less than 19 $\mearth$, while the synthetic planets that agree with the observations in Fig.  \ref{fig:INTRObetapicb} have much more massive cores exceeding $\sim$100 $\mearth$. For identical core masses, the two models agree well. This insight from the population synthesis led to a systematic dedicated study of the dependency of the post-formation entropy and luminosity on the core mass \citep{mordasini2013}. This is therefore another example how global models feed back into specialized ones. It was found that the post-formation luminosity of massive giant planets is very sensitive to the core mass due to a self-amplifying mechanism (see \citealt{mordasini2013} and \citealt{bodenheimerdangelo2013} for details). 

It is currently unclear whether such massive cores can  actually form, especially at large orbital distances. Several factors play a role, like  {for example} the settling of dissolved planetesimal material towards  the center of the planet \citep{iaroslavitzpodolak2007} and the timescale of core accretion which might get speeded up if mainly small bodies are accreted \citep[see, e.g.,][]{dodson-robinsonveras2009,ormelklahr2010,krattermurray-clay2011,lambrechtsjohansen2012}. At least for some close-in transiting giant planets, very large core masses ($\gtrsim100\mearth$) have been inferred  {as mentioned} from the observed mass-radius relationship \citep{millerfortney2011,deleuilbonomo2012}.

\section{Summary}	
We have reviewed global planet formation models that are used in population synthesis calculations. Such global models predict the properties of a planetary system based on the properties of a protoplanetary disk. They therefore establish a link between two classes of observable astrophysical objects which are on one hand the protoplanetary disks as the initial conditions for the planet formation process and on the other hand the extrasolar planets that are the final outcome of this process.

Global models have mainly been used in statistical studies. Such studies are a young approach that helps to improve the theory of planet formation by the comparison of theoretical results and statistical observational constraints provided by the entire population of extrasolar planets. With this approach, the global effects of many different physical mechanisms occurring during planet formation and evolution can be assessed, and (often strongly simplified) theoretical descriptions of these  {processes} derived from specialized models can be tested against observations. 

The first group of extrasolar planets that was known in sufficient numbers for statistical comparisons were giant planets detected by the radial velocity method. Population synthesis calculations therefore initially concentrated on studying the mass and semimajor axis distribution of this type of planets \citep{idalin2004a,mordasinialibert2009}. After this first phase of extrasolar planet detection, recent observational progress now also provides a first geophysical characterization of a growing number of planets outside of the Solar System (Sect. \ref{sect:fromformationtoevolution}). This has led to additional observational constraints for global models like the planetary mass-radius relationship (Fig. \ref{fig:INTROmrobs}) or the  intrinsic luminosities of young giant  planets  (Section \ref{sect:INTROdirectimaging}). 

In order to yield synthetic populations that can be directly compared with all these different techniques, global models unite in one point the essential results of a significant number of specialized sub-models that describe one specific physical mechanism. The global formation and evolution model mainly discussed in this paper  \citep{alibertmordasini2005,mordasinialibert2012a,alibertcarron2013}, for example consists of eleven sub-models (Fig. \ref{fig:INTROschema}) addressing the following  {aspects}: (1) the vertical structure of the protoplanetary disk. (2) the radial structure of the protoplanetary disk. (3) the disk of solids (planetesimals). (4) the core accretion rate of the protoplanet. (5) the planetary gas envelope. (6) the atmosphere of the (proto)planet. (7) the infall of planetesimals into the protoplanet's envelope. (8) the internal structure of the solid core. (9) the atmospheric escape (envelope evaporation) during evolution. (10) the orbital migration due to tidal interaction with the protoplanetary gas disk  {and finally} (11) the gravitational interaction between the protoplanets. There are therefore three different classes of sub-models, namely those that describe the protoplanetary disk (1-3), those that describe one (proto)planet (4-9), and those that describe the interactions (10-11). In this paper, we have presented  detailed descriptions of the physics included in these different sub-models and addressed their  limitations and possible future improvements. 

Due to their nature as meta models, the global models discussed in this review depend directly on the results of many different specialized models, and therefore  on the development of the entire field of planet formation theory. There are important uncertainties in this theory regarding even key aspects, therefore it is likely that the global models presented here will in future undergo significant modifications. These aspects include (1) the formation of planetesimals and the resulting ``initial'' distribution of solids in the disk. (2) the accretion of the solid core. (3) the opacity in the protoplanetary atmosphere and the associated gas accretion timescale. (4) the efficiency of orbital migration (which is still too rapid even with non-isothermal migration, see  Fig. \ref{fig:INTROtracks}) and (5) the magnitude of gas accretion in the runaway phase. The later two points could be addressed with 2D hydrodynamic simulations which can now be run over long timescales (\citealt{zhunelson2011}) instead of 1D disk models.  It is also possible that at some point it becomes necessary to abandon the simplifications that planetary systems form in isolation, because the gravitational interaction with other stars in young stellar clusters could be important \citep{malmbergdeangeli2007}. 

The description of planetary evolution in the global models should in future include better atmospheric models and address the effects of heavy element settling, core erosion in giant planets, and eventually the formation of secondary atmospheres based on the composition acquired during the planets' formation for low mass planets. Future global models of planet formation and evolution should therefore also include  better descriptions of condensation and disk chemistry, so that the resulting composition of the planets can be predicted in a more detailed and self-consistent way \citep{thiabaudmarbeuf2013}.  On a longer timescale, this should make it possible to predict the habitability of a planet based on its formation.
 
Despite the current limitations, when used in planetary population synthesis calculations, global models can already now yield many testable predictions for the major observational techniques. This is important in a time where many surveys both from space and ground yield or will soon yield large amounts of additional data on both the global statistics and the detailed physical characteristics of extrasolar planets. Seeking for the theoretical models that best explain these combined data sets will be a promising approach towards a better understanding of planet formation and evolution. 

\acknowledgements{We thank Micka\"el Bonnefoy, Gabriel Marleau, Thomas Henning, Chris Ormel, and Willy Benz for interesting discussions.  This work was supported in part by the Swiss National Science Foundation and the European Research Council under grant 239605. CM thanks the Max Planck Society for the Reimar-L\"ust Fellowship.  {We thank an anonymous referee for comments that helped to improve the manuscript.}}

\bibliographystyle{aa} % style aa.bst \bibliography{Yourfile} % your references Yourfile.bib \end{document}
\bibliography{biball2013new}

\end{document}